\documentclass[twocolumn,aps,superscriptaddress,floatfix,tightenlines]{revtex4-1}
\def\onlinecite{\cite}

\usepackage{graphicx}
\usepackage{latexsym,savesym,amsmath,amssymb,amsfonts,bm,euscript}
\usepackage{color}
\usepackage{hyperref}
\usepackage{subfigure}
\usepackage{wasysym}

\def\kappaET{{$\kappa$-(BEDT-TTF)$_{2}$Cu$_{2}$(CN)$_{3}$}}
\def\dmit{{EtMe$_{3}$Sb[Pd(dmit)$_{2}$]$_{2}$}}

\newcommand{\Lag}{\mathcal{L}}
\newcommand{\Ham}{\mathcal{H}}

\def\up{\uparrow}
\def\dn{\downarrow}
\begin{document}

\title{A continuous Mott transition between a metal and a quantum spin liquid}

\author{Ryan V.~Mishmash}
\affiliation{Department of Physics, University of California, Santa Barbara, California 93106, USA}
\affiliation{Walter Burke Institute for Theoretical Physics, California Institute of Technology, Pasadena, California 91125, USA}

\author{Iv\'an Gonz\'alez}
\affiliation{Centro de Supercomputaci\'on de Galicia, Avda.~de~Vigo~s/n, E-15705 Santiago de Compostela, Spain}

\author{Roger G. Melko}
\affiliation{Department of Physics and Astronomy, University of Waterloo, Ontario, N2L 3G1, Canada}
\affiliation{Perimeter Institute for Theoretical Physics, Waterloo, Ontario, N2L 2Y5, Canada} 

\author{Olexei I.~Motrunich}
\affiliation{Department of Physics, California Institute of Technology, Pasadena, California 91125, USA}

\author{Matthew P.~A.~Fisher}
\affiliation{Department of Physics, University of California, Santa Barbara, California 93106, USA}

\date{\today}

\begin{abstract}
More than half a century after first being proposed by Sir Nevill Mott, the deceptively simple question of whether
the interaction-driven electronic metal-insulator transition may be continuous remains enigmatic.  
Recent experiments on two-dimensional materials suggest that when the
insulator is a quantum spin liquid, lack of magnetic
long-range order on the insulating side may cause the transition to be continuous,
or only very weakly first order.
Motivated by this, we study a half-filled extended Hubbard model on a triangular lattice strip geometry.
We argue, through use of large-scale numerical simulations and analytical bosonization,
that this model harbors a continuous (Kosterlitz-Thouless-like) quantum phase transition
between a metal and a gapless spin liquid characterized by a spinon Fermi surface, i.e., a ``spinon metal''.
These results may provide a rare insight into the development of Mott criticality in  
strongly interacting two-dimensional materials and represent one of the first
numerical demonstrations of a Mott insulating quantum spin liquid phase in a genuinely electronic
microscopic model.
%elucidate a mechanism by which spin-liquid phases are stabilized in the vicinity of such transitions.
\end{abstract}

\maketitle

\section{\uppercase{Introduction}}

Strongly correlated electronic systems may have insulating phases 
that originate entirely from electron-electron interactions.  These insulators, and 
their phase transitions to metallic phases have a long history 
reaching back into the pioneering work
of Mott~\cite{Mott90_MITs, Imada:1998er}.  However, despite decades of study, 
metal-insulator transitions driven by strong correlations---Mott's namesake---remain
rather poorly understood.  Central to this
difficulty is the fact that Mott transitions exhibit strong quantum fluctuations,
which can inherit correlations from both the adjacent metallic and insulating
phases.  Thus, the nature of the Mott transition
may depend crucially on the properties of each of these phases.

Conventional insulating phases, such as those with magnetic long-range order,
appear to predominantly give rise to first-order Mott transitions, as has been
observed in a number of experimental systems in the past~\cite{Mott75_PRB_11_4383,
Georges03_Science_302_89, Georges03_PRL_91_016401,
Kanoda04_PRB_69_064511, Kanoda05_Nature_436_534}.  The reason for first-order
behavior is simple:  The properties of \emph{both} the spin and charge sectors
change qualitatively at the transition, the former developing magnetic
long-range order and the latter localizing to form an insulating state.  In
contrast, systems that harbor unconventional, exotic insulating phases showing
no symmetry breaking down to zero temperature---so-called quantum spin liquids~\cite{Anderson:1973,
Anderson:1987gf, Lee:2006de, Balents:2010ds}---offer a
promising playground for finding the long-sought-after continuous Mott
transition.  For example, one beautiful possiblity is that the spin sector on
the insulating side may be described by a spinon Fermi surface coupled to a
U(1) gauge field~\cite{Lee:2005ew} (the so-called ``spin Bose metal''~\cite{Sheng:2009up},
hereafter referred to as simply the ``spinon metal'').
In this case, the behavior of the spin correlations
would be qualitatively unchanged~\cite{Motrunich:2005df} upon crossing the
transition, making the nature of the transition determined entirely by the
charge sector.  Thus, as proposed in Refs.~\onlinecite{Georges04_PRB_70_035114,
Senthil:2008ki, Mross:2011cq}, perhaps the electronic Mott transition
in $d$ spatial dimensions can be in the ($d+1$)D XY universality class,
the same as obtained for bosons~\cite{Fisher89_PRB_40_546}!

\def\ladderCaption
{
\textbf{Schematic of the half-filled extended Hubbard model on the two-leg triangular strip and its phase diagram.}
Top:  Our electronic model contains electron hoppings $t$ and $t'$ in addition to repulsive Hubbard interactions
up to fourth neighbor [see Eqs.~(\ref{eq:hamiltonian})-(\ref{eq:modelV})].
As shown, we view the two-leg triangular strip as a 1D chain and attack the problem with DMRG and bosonization.
Bottom:  The phase diagram of our model as a function $U/t$ for the chosen characteristic parameters (see text).
}

\begin{figure}[b]
\vspace{-0.15in}
\includegraphics[width=1.0\columnwidth]{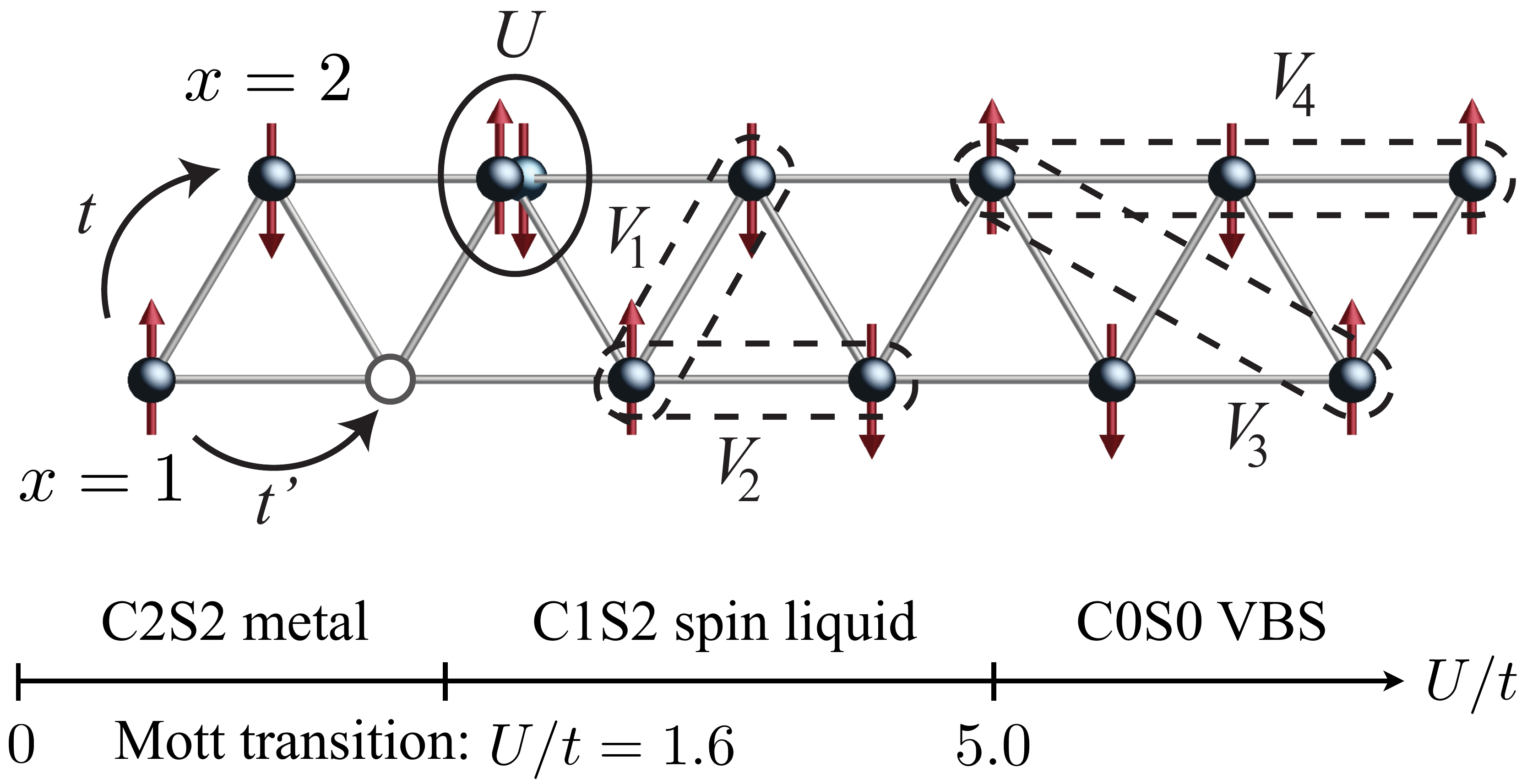}
\vspace{-0.15in}
\caption{\ladderCaption}
\label{fig:ladder}
\end{figure}

Fortunately, this sort of physics is more than just a theorist's dream, as recently
several experimental groups have found strong evidence for spin-liquid behavior proximate to a Mott transition in 
two separate quasi-two-dimensional triangular lattice organic materials.
In 2003, a putative spin-liquid phase in $\kappa$-(BEDT-TTF)$_{2}$Cu$_{2}$(CN)$_{3}$ 
was discovered~\cite{Shimizu:2003gq}, which is insulating at ambient pressure with no apparent long-range order but
can indeed be driven metallic by application of moderate pressure~\cite{Kurosaki:2005ji}. 
More recently, Itou \emph{et al.}~\cite{Itou:2008} found a spin-liquid candidate in
EtMe$_{3}$Sb[Pd(dmit)$_{2}$]$_{2}$.  Further experiments
indicated the existence of highly mobile gapless spin excitations in both compounds~\cite{Yamashita:2008ca,
Yamashita04062010}, although the precise nature of the spin excitations in 
$\kappa$-(BEDT-TTF)$_{2}$Cu$_{2}$(CN)$_{3}$ at the lowest temperatures
is still highly controversial~\cite{Yamashita:2008cc}.
These findings suggest that the spinon metal is likely a good starting point for
understanding the spin-liquid behavior observed in these two materials~\cite{Lee:2005ew, Motrunich:2005df}.
In addition, the pressure-induced Mott transition from the metal to the spin liquid is observed to be either
only very weakly first order~\cite{Kurosaki:2005ji},
or perhaps even continuous~\cite{Witczak-Krempa12_PRB_86_245102, Kanoda12_fragnetsConfTalk,
Kanoda15_NaturePhys_11_221}.

Motivated by these experiments, we consider a model of interacting
electrons on a half-filled triangular lattice ``strip'' geometry (see Fig.~\ref{fig:ladder}),
which we solve using large-scale density matrix renormalization group (DMRG) calculations.
By increasing the strength of the repulsive electron-electron interactions, we drive
the ground state of the system from a metallic Fermi liquid-like phase
to an insulating phase identified as the electronic spinon metal~\cite{Lee:2005ew, Motrunich10_PRB_81_045105}
via an intervening continuous Kosterlitz-Thouless-like quantum phase transition.
Our realization of this spin liquid phase constitutes perhaps the first numerical
demonstration of a Mott insulating quantum spin liquid in an interacting microscopic model involving
itinerant electrons that is beyond the strictly one-dimensional (one-band) limit~\cite{Giamarchi}.
Furthermore, we are able to characterize this exotic phase in a very thorough fashion.
Further increasing the electron interactions eventually drives the system into
a spin-gapped valence bond solid (VBS) insulator---the phase realized by
the effective Heisenberg spin model that our half-filled electronic model approaches at strong repulsion.
Our calculations thus represent a direct quasi-one-dimensional (quasi-1D) analog of tuning a two-dimensional (2D)
half-filled Hubbard-type model from a metal to a quantum spin liquid
to a conventional ordered phase via increasing overall electron repulsion~\cite{Kawakami09_PRL_103_036401,
Lauchli10_PRL_105_267204, McKenzie13_PRL_110_206402, Meng10_Nature_464_847},
a result with clear potential relevance to the Mott physics observed
in the organic spin liquid materials~\cite{Kurosaki:2005ji, Kato06_DMITmott, Kanoda15_NaturePhys_11_221}.

\section{\uppercase{Extended Hubbard model on the two-leg triangular strip}}

The most appropriate microscopic model for the triangular-lattice organic materials
is a Hamiltonian consisting of electron hopping plus moderately strong,
possibly extended~\cite{Imada09_JPSJ_78_083710, Hotta14_PRB_89_045102}, Coulomb repulsion.
As is well-known from some 30 years of research on the high-temperature cuprate superconductors~\cite{Lee:2006de},
such a model does not succumb easily to either exact analytical field theory
nor direct numerical simulations in two dimensions due to the fermionic ``sign problem''.

Recently, some of us have proposed a novel approach to the 2D limit of such models
through a sequence of studies on quasi-1D ladder geometries, which have the 
significant advantage that they can be solved exactly with DMRG~\cite{White:1992ie,
White93_PRB_48_10345, Schollwock:2005wz}.
Sheng \emph{et al.}~used this line of attack to extensively study an effective spin model
appropriate for the ``weak'' Mott insulating regime of the organic materials~\cite{Motrunich:2005df,
Lauchli10_PRL_105_267204} and indeed found exceptionally strong evidence
that quasi-1D descendants of the spinon metal exist as the ground state over a large region
of the phase diagram~\cite{Sheng:2009up, Block:2011bv}.
The low-energy degrees of freedom of this exotic spin liquid are modeled as mobile and charge-neutral
spin-1/2 fermionic spinons coupled to a U(1) gauge field.  In 2D, these gapless spinons give rise to
a spin structure factor with power-law singularities residing on an entire ``Bose surface'' in momentum space.
However, in quasi-1D the Bose surface is reduced to a set of points,
so that quasi-1D descendants of the 2D spin liquid are dramatically recognizable on ladders,
making the quasi-1D approach very fruitful~\cite{Sheng:2009up, Block:2011bv,
Mishmash11_PRB_84_245127, Jiang13_Nature_493_39}.

Inspired by these recent developments and restricting ourselves to the two-leg triangular strip
for numerical tractability, we consider the following extended Hubbard model (see Fig.~\ref{fig:ladder}):
\begin{align}
    H=&-\sum_{x,\alpha}\left[t\,c^{\dagger}_{\alpha}(x)c_{\alpha}(x+1) + t'c^{\dagger}_{\alpha}(x)c_{\alpha}(x+2)+\mathrm{H.c.}\right]\nonumber\\
    &+\frac{1}{2}\sum_{x,x'}V(x-x')n(x)n(x'), \label{eq:hamiltonian}
\end{align}
where $c_{\alpha}(x)$ destroys an electron at site $x$ with spin ${\alpha=\,\uparrow,\downarrow}$\,,
$n(x)\equiv\sum_\alpha c^{\dagger}_{\alpha}(x)c_{\alpha}(x)$ is the electron number operator,
and we take the system to be half-filled with one electron per site.

\def\dispersionCaption
{
\textbf{Electron/spinon bands on the two-leg triangular strip.}
In the noninteracting $U/t=0$ limit, the ground state of our model for $t'/t>0.5$ consists
of two disconnected Fermi seas (bands) with Fermi points as labeled above.
On the other hand, the insulating two-band spinon metal can be modeled,
in a pure spin system, by Gutzwiller projecting the same band structure (see Ref.~\onlinecite{Sheng:2009up}).
Here, we realize a continuous Mott transition between these two phases driven at strong interactions
by an eight-fermion umklapp term which scatters both spin-up and spin-down electrons across each Fermi sea
(black arrows).
}

\begin{figure}[b]
\vspace{0.07in}
\includegraphics[width=1.0\columnwidth]{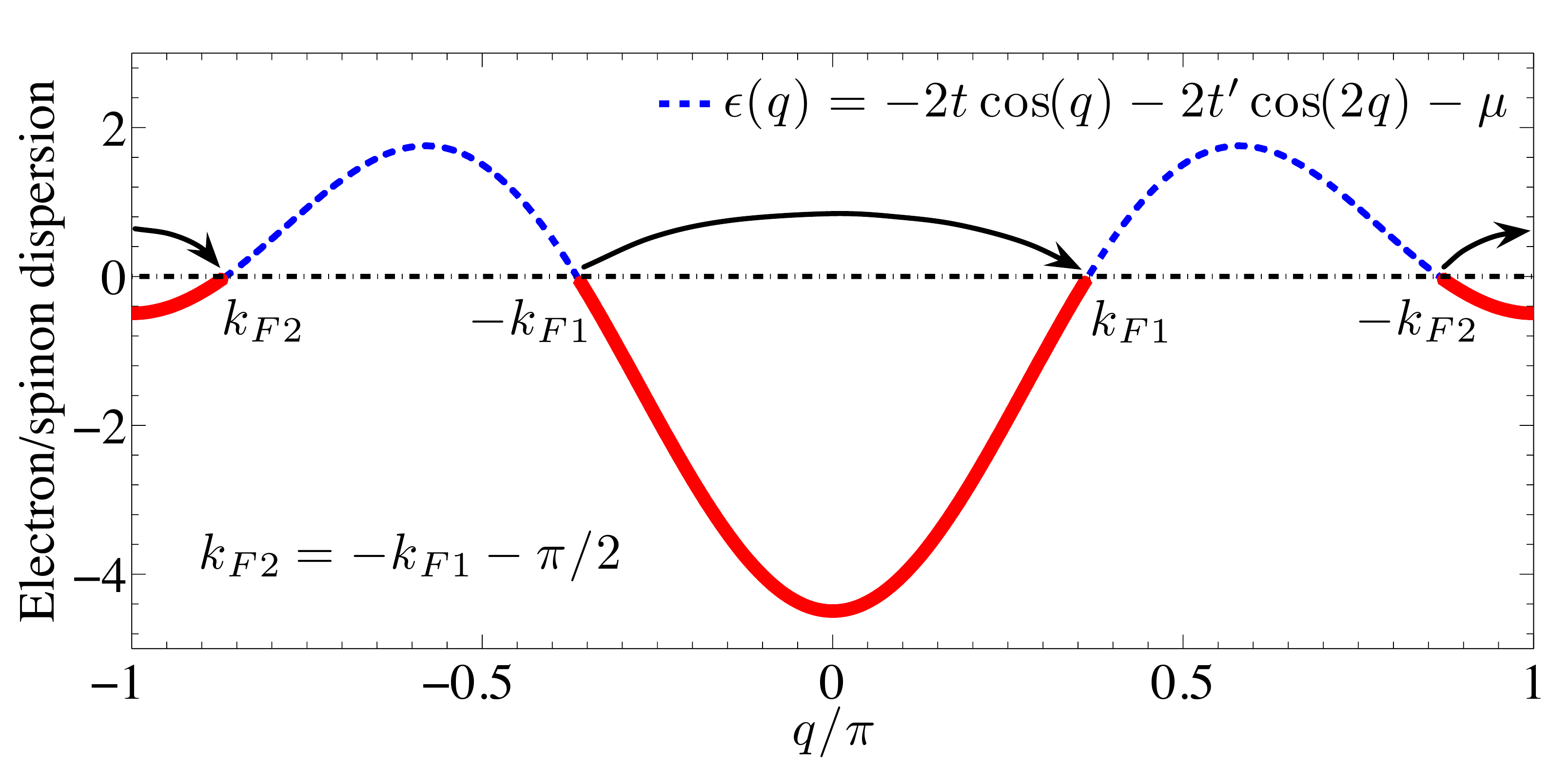}
\caption{\dispersionCaption}
\label{fig:dispersion}
\end{figure}

In the usual on-site Hubbard model, we would have $V(x-x')=U\delta_{x,x'}$.
However, inspired by the results of Ref.~\onlinecite{Motrunich10_PRB_81_045105},
we allow for longer-ranged repulsion in our Hamiltonian.
For concreteness, we take the following model potential:
\begin{align}
V(x-x') = 
\begin{cases}
\,U &, \hspace{0.5cm} |x-x'|=0 \\
\,\kappa U e^{-\gamma |x-x'|} &, \hspace{0.5cm} 1 \leq |x-x'| \leq 4 \\
0 &, \hspace{0.5cm} |x-x'| > 4\,.
\end{cases}
\label{eq:modelV}
\end{align}
The reasoning for considering such longer-ranged repulsion in the model Hamiltonian is twofold.
First, such terms are well-motivated by recent \emph{ab initio} calculations~\cite{Imada09_JPSJ_78_083710, Hotta14_PRB_89_045102},
which indicate a substantial long-ranged tail in the effective screened Coulomb repulsion appropriate for \kappaET.
Second, on the two-leg ladder, such terms fight the spin-gap tendencies
present in the metallic phase of the $t$-$t'$-$U$ Hubbard model
(i.e., our model with $\kappa=0$; see, for example,
Refs.~\onlinecite{Balents96_PRB_53_12133, Gros01_PRB_64_113106, Japaridze07_PRB_76_115118}),
thus at least allowing for the possibility of a direct, continuous transition
between a spin gapless two-band metal and two-band spinon metal spin liquid.
Guided by the weak and intermediate coupling analysis of Ref.~\onlinecite{Motrunich10_PRB_81_045105},
in what follows we choose characteristic parameters $t'/t=0.8$, $\kappa=0.5$, and $\gamma=0.2$,
leaving the single dimensionless ratio $U/t$ to control the overall strength of electron repulsion.

\section{\uppercase{Mott metal-insulator transition and realization of the electronic spinon metal}}

We first sketch the low-energy effective theory describing the putative metal to spinon metal transition
and then present strong numerical evidence that this exotic scenario is indeed realized. %in Eq.~(\ref{eq:hamiltonian}).
In the absence of interactions ($U/t=0$), our model for $t'/t>0.5$ simply describes two bands of noninteracting spinful electrons
(see Fig.~\ref{fig:dispersion}).  Importantly, the weak-coupling analysis of Ref.~\onlinecite{Motrunich10_PRB_81_045105}
indicates that this spin gapless two-band metallic state---so-called C2S2 in the literature, where
C$\alpha$S$\beta$ denotes a Luttinger liquid with $\alpha$ gapless charge modes
and $\beta$ gapless spin modes~\cite{Balents96_PRB_53_12133}---is stable
in our extended Hubbard model, Eqs.~(\ref{eq:hamiltonian})-(\ref{eq:modelV}),
in the presence of infinitesimal $U/t$.
At half-filling, there is an allowed \emph{eight-fermion} umklapp term in our two-band system
(see Fig.~\ref{fig:dispersion}).  Bosonizing (see, e.g.,
Refs.~\onlinecite{Haldane81_PRL_47_1840, Lin98_PRB_58_1794, Fisher99_LesHouches, Giamarchi})
this interaction gives
\begin{equation}
\Ham_8=2u\cos(4\theta_{\rho+}),
\label{eq:H8main}
\end{equation}
where $\theta_{\rho+}$ is the density field for the overall charge mode,
i.e., $\delta n(x) = 2\partial_x\theta_{\rho+}/\pi$ is the coarse-grained electron density.
Assuming the C2S2 metal is stable against opening of a spin gap~\cite{Motrunich10_PRB_81_045105},
then the fixed-point Lagrangian $\mathcal{L}_\mathrm{C2S2}$ involves four gapless bosonic modes,
one being $\theta_{\rho+}$
(see Appendix~\ref{sec:bosonization} and Ref.~\onlinecite{Motrunich10_PRB_81_045105} for details).
For free electrons, the scaling dimension of the eight-fermion umklapp term is $\Delta[\Ham_8]=4>2$,
so that $\Ham_8$ is strongly irrelevant at weak coupling.
However, increasing $U/t$ in our microscopic model will feed into ``stiffening'' $\theta_{\rho+}$
in $\mathcal{L}_\mathrm{C2S2}$, thus decreasing $\Delta[\Ham_8]$.
Eventually $\Delta[\Ham_8]=2$, beyond which the umklapp is relevant so that $u$ grows
at long scales pinning $\theta_{\rho+}$ into one of the minima of the cosine potential in $\Ham_8$.
The resulting phase is a remarkable C1S2 Luttinger liquid,
which is precisely the electronic spinon metal~\cite{Sheng:2009up},
The remaining ``charge mode'' does not transport charge along the ladder
but rather represents local current loop fluctuations;
it encodes long-wavelength fluctuations of the spin chirality as discussed in Ref.~\onlinecite{Sheng:2009up}.

The critical theory describing the C2S2$\rightarrow$C1S2 metal-insulator transition
is a sine-Gordon-like theory~\cite{Amit80_JPhysA_13_585},
with a technical complication arising because $\theta_{\rho+}$ is coupled to the ``relative charge'' field $\theta_{\rho-}$
in $\mathcal{L}_\mathrm{C2S2}$ (see Appendix~\ref{sec:bosonization}).
Nonetheless, the transition is still Kosterlitz-Thouless-like~\cite{Weber88_PRB_37_5986} [($1+1$)D XY]
and represents a direct, nontrivial two-leg analog of the ($2+1$)D scenario
recently proposed by Senthil~\cite{Senthil:2008ki, Mross:2011cq}.

We now present our numerical results, giving strong evidence that the above scenario is actually realized.
To numerically characterize the system, we focus on four main quantities:
the density structure factor $\langle\delta n_{q}\delta n_{-q}\rangle$,
the spin structure factor $\langle\mathbf{S}_{q}\cdot\mathbf{S}_{-q}\rangle$,
the dimer structure factor $\langle\mathcal{B}_q\mathcal{B}_{-q} \rangle$,
and the electron momentum distribution function $\langle c_{q\alpha}^\dagger c_{q\alpha}\rangle$,
where $\delta n_{q}$, $\mathbf{S}_{q}$, $\mathcal{B}_{q}$, and $c_{q\alpha}$
are the Fourier transforms of the local operators
$\delta n(x)\equiv n(x)-\langle n(x)\rangle$,
$\mathbf{S}(x)\equiv\frac{1}{2}\sum_{\alpha,\beta}c^{\dagger}_{\alpha}(x)\bm{\sigma}_{\alpha\beta}c_{\beta}(x)$,
$\mathcal{B}(x)\equiv\mathbf{S}(x)\cdot\mathbf{S}(x+1)$,
and $c_{\alpha}(x)$, respectively.
In the data presented here, we consider systems up to $L=96$ sites with periodic boundary conditions.
(See Appendix~\ref{sec:observables_def} for all details,
including discussion of the chosen boundary conditions.)

We focus first on the density (charge) structure factor $\langle\delta n_{q}\delta n_{-q}\rangle$.
A crucial aspect of  $\langle\delta n_{q}\delta n_{-q}\rangle$ lies in its ability
to distinguish metallic from insulating behavior at small wavevectors $q$.
For a metallic state, we expect $\langle\delta n_{q}\delta n_{-q}\rangle\sim |q|$ for $q\sim0$.
Specifically, for the two-band C2S2 metal, the slope of $\langle\delta n_{q}\delta n_{-q}\rangle$
at $q=0$ is related to the ``Luttinger parameter'' $g_{\rho+}$ for the overall charge mode $\theta_{\rho+}$:
\begin{equation}
\langle\delta n_{q}\delta n_{-q}\rangle= 2g_{\rho+} |q|/\pi~~\mathrm{as}~~q\rightarrow0.
\label{eq:dndnqSlope}
\end{equation} 
Importantly, the quantity $g_{\rho+}$ as determined from Eq.~(\ref{eq:dndnqSlope}) gives a direct
measure of the scaling dimension of $\Ham_8$:  $\Delta[\Ham_8]=4g_{\rho+}$ (see Appendix~\ref{sec:observables_theory}).
Once $\Delta[\Ham_8]<2$ [corresponding to measured $g_{\rho+}<1/2$ in Eq.~(\ref{eq:dndnqSlope})],
then the umklapp is relevant, and the system is necessarily insulating.
We then expect $g_{\rho+}\rightarrow0$ at long scales so that
$\langle\delta n_{q}\delta n_{-q}\rangle$ becomes quadratic at small $q$:
$\langle\delta n_{q}\delta n_{-q}\rangle\sim q^2$ in the Mott insulator.

In Fig.~\ref{fig:charges}, we show a series of density structure factor measurements ranging from
the noninteracting limit at $U/t=0$ to deep in the Mott insulating phase at $U/t=7.0$.
In the inset, we show estimates of $g_{\rho+}$ by plotting $\langle\delta n_{q}\delta n_{-q}\rangle/(2|q|/\pi)$
[see Eq.~(\ref{eq:dndnqSlope})].  Based on the above arguments, we see that the Mott transition occurs
near a critical value of $U/t=1.6$ where $g_{\rho+}$ drops below 1/2.  Note, however, that for these system sizes
$\langle\delta n_{q}\delta n_{-q}\rangle$ still appears linear in $q$ until much larger overall repulsion, i.e., $U/t\simeq5.0$.
Still, we argue that the system becomes insulating at $U/t=1.6$, as this is where $\Ham_8$ is determined to be
relevant based on the measurement of $g_{\rho+}$.  That is, we, rather remarkably, have an insulating state with a
charge correlation length comparable to our system size ($L=96$) for $1.6\lesssim U/t \lesssim 5.0$.
Indeed, such large correlation lengths are expected in the weak Mott insulating spinon metal,
which we now argue is precisely the phase realized immediately on the insulating side of our model.
(For more discussion on the finite-size behavior of $g_{\rho+}$,
we refer the reader to Appendix~\ref{sec:observables_theory}.)

\def\chargesCaption
{
\textbf{Density structure factor: Locating the Mott transition and power-law Friedel oscillations in a Mott insulator.}
Measurements of the density structure factor, $\langle\delta n_{q}\delta n_{-q}\rangle$,
allow us to locate the Mott transition near $U/t=1.6$ (black curve with $\ast$ symbols).
The onset of the Mott transition occurs when the overall charge Luttinger parameter $g_{\rho+}$
drops below 1/2.  We measure $g_{\rho+}$ via the slope of $\langle\delta n_{q}\delta n_{-q}\rangle$ at $q=0$,
as shown in the inset [see Eq.~(\ref{eq:dndnqSlope})].  For $U/t>1.6$, the system is insulating, yet displays power-law
singularities in $\langle\delta n_{q}\delta n_{-q}\rangle$ at finite wavevectors~\cite{Mross11_PRB_84_041102}
(see black $\star$ and hexagram symbols).
Data correspond to a system of length $L=96$
with $U/t=0, 0.4, 0.8, 1.2, 1.6, 2.0, 3.0, 4.0, 5.0, 6.0, 7.0$ (from top to bottom, blue to red).
%All raw data in our manuscript is available online at Ref.~\onlinecite{GithubRepo_data}.
}

\begin{figure}[t]
\includegraphics[width=1.0\columnwidth]{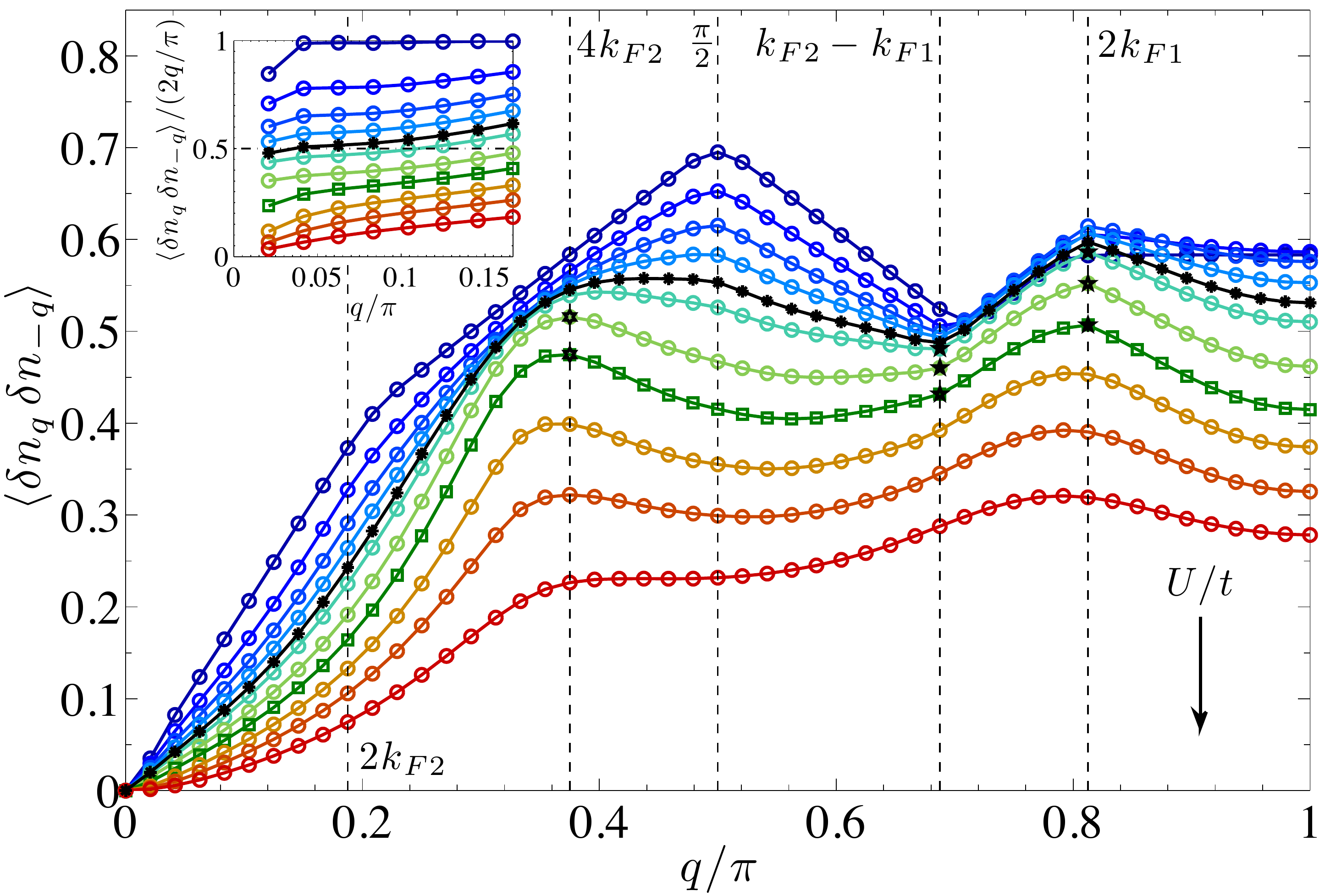} 
\caption{\chargesCaption}
\label{fig:charges}
\end{figure}

To this end, we now turn to the spin structure factor $\langle\mathbf{S}_{q}\cdot\mathbf{S}_{-q}\rangle$
in Fig.~\ref{fig:spins}.  In the noninteracting limit $U/t=0$, we have familiar singularities at wavevectors
$q=2k_{F1}, 2k_{F2}, \pi/2, k_{F2}-k_{F1}$ originating from various ``$2k_F$'' processes
in our two-band system (see Fig.~\ref{fig:dispersion}).
These singularities are simple slope discontinuities, i.e., the scaling dimension for the spin
operator at each wavevector is unity as guaranteed by Wick's theorem.
As we enter the putative interacting C2S2 metal by turning on finite $U/t$,
the scaling dimensions at wavevectors $2k_{F1}, 2k_{F2}, \pi/2, k_{F2}-k_{F1}$
are renormalized slightly but remain near unity.

Near the Mott transition value $U/t=1.6$ as determined from $\langle\delta n_{q}\delta n_{-q}\rangle$ above,
we observe the remarkable result that the singular features in $\langle\mathbf{S}_{q}\cdot\mathbf{S}_{-q}\rangle$
all survive, and those at $q=2k_{F1}, 2k_{F2}, \pi/2$ are actually \emph{enhanced} upon entering the insulating phase.
Indeed, these are characteristic signatures of the spinon metal.
(In Figs.~\ref{fig:charges}-\ref{fig:dimers}, we display characteristic C1S2 spinon metal data at $U/t=4.0$
with distinctive dark green square symbols.)
First, the singular features in $\langle\mathbf{S}_{q}\cdot\mathbf{S}_{-q}\rangle$ still correspond to
the same ``$2k_F$'' processes as in the metallic phase, but with the charge gapped they now correspond
to \emph{spinon} transfers across the Fermi sea.
Second, in the spinon metal, we indeed expect the scaling
dimensions of the spin operator at wavevectors $2k_{F1}, 2k_{F2}, \pi/2$ to be decreased
(singularities enhanced) from their mean-field values~\cite{Sheng:2009up}.
This enhancement can be understood clearly within the bosonization framework.
Specifically, when written in terms of bosonized fields, the slowly varying part of the spin operator
at wavevectors $Q=2k_{F1}, 2k_{F2}, \pi/2$ contains directly the field $\theta_{\rho+}$, i.e.,
$\mathbf{S}_Q\sim e^{\pm i\theta_{\rho+}}(\cdots)$---see Appendix~\ref{sec:observables_theory}
and Ref.~\onlinecite{Sheng:2009up}.  Thus, pinning of $\theta_{\rho+}$ at the Mott transition
reduces the fluctuating content of the spin operator at these wavevectors,
which in turn reduces the scaling dimensions and, ultimately, enhances the structure factor singularities.
This enhancement is actually a (1+1)D realization of ``Amperean'' attraction between a spinon ``particle'' and ``hole''
moving in opposite directions~\cite{Lee:2006de, Sheng:2009up}.

\def\spinsCaption
{
\textbf{Spin structure factor: Watching electrons evolve into spinons.}
Measurements of the spin structure factor, $\langle\mathbf{S}_{q}\cdot\mathbf{S}_{-q}\rangle$,
strongly point toward the presence of gapless spin excitations in both the metal
and putative spinon metal immediately after the Mott transition at $U/t=1.6$ (black curve with $\ast$ symbols).
Gapless spin excitations are characterized by $\langle\mathbf{S}_{q}\cdot\mathbf{S}_{-q}\rangle\sim|q|$ as $q\to0$,
and, as shown in the top inset, the opening of a spin gap occurs only for $U/t\gtrsim5.0$,
at which point the system dimerizes.
The ``$2k_{F}$'' features of the two electron bands in the metallic phase are inherited by the two spinon bands
in the spinon metal, and, as highlighted in the bottom inset for $q=2k_{F2}$, they are actually enhanced.
Data correspond to the same $U/t$ values and color scheme as in Fig.~\ref{fig:charges}.
}

\begin{figure}[t]
\includegraphics[width=1.0\columnwidth]{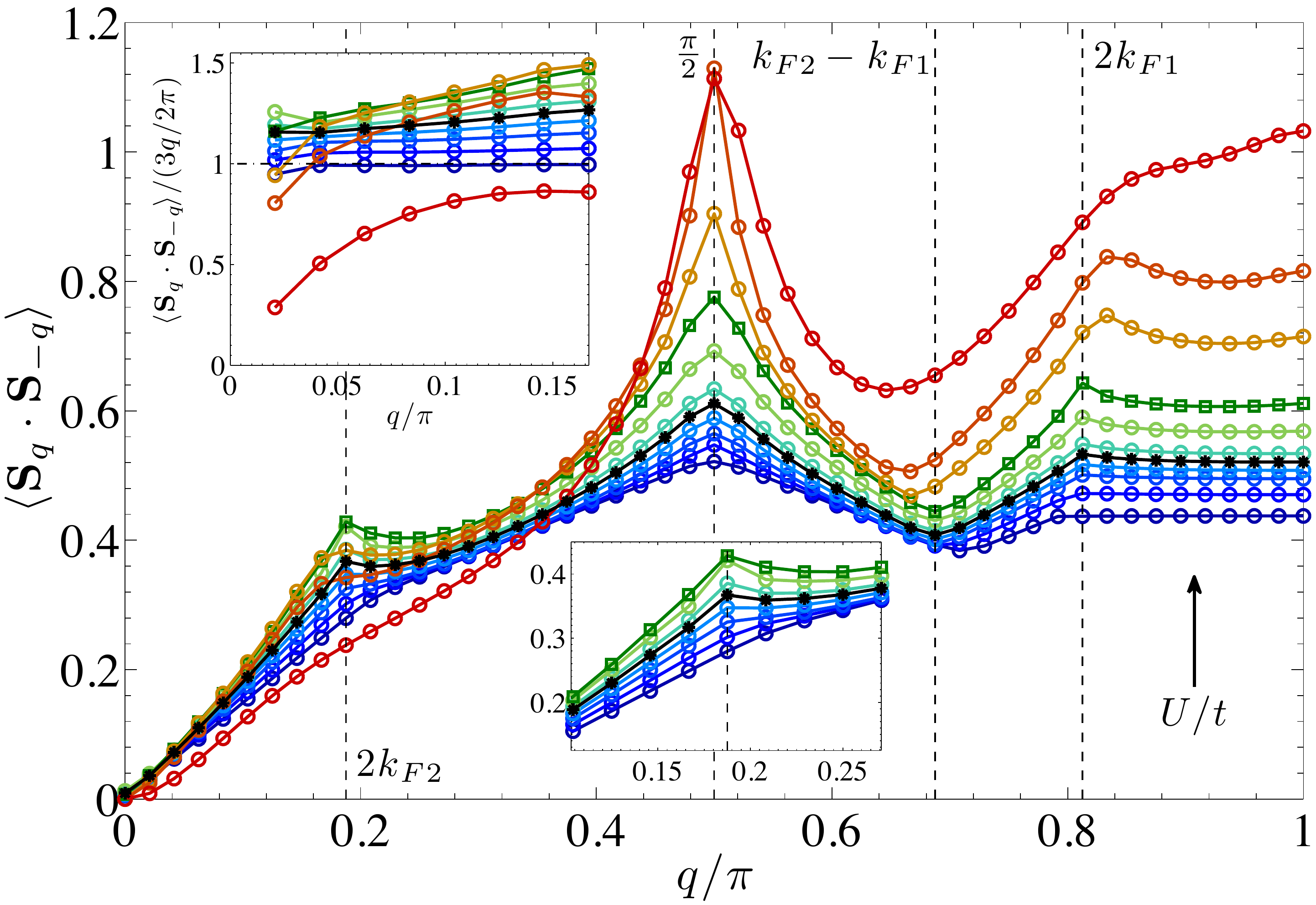} 
\caption{\spinsCaption}
\label{fig:spins}
\end{figure}

In the density structure factor measurements of Fig.~\ref{fig:charges},
we also have singular features at the ``$2k_F$'' wavevectors $q=2k_{F1}, 2k_{F2}, \pi/2, k_{F2}-k_{F1}$
within the metallic phase, and in fact in the noninteracting $U/t=0$ limit, the density and spin structure factors
as defined are identical:  $\langle\delta n_{q}\delta n_{-q}\rangle = \frac{4}{3} \langle\mathbf{S}_{q}\cdot\mathbf{S}_{-q}\rangle$.
In the interacting C2S2 metal, the features at $q=2k_{F1}, \pi/2, k_{F2}-k_{F1}$ are still clearly visible.
In fact, some of these features survive even upon entering the putative insulating spinon metal
and remain until $U/t\simeq4.0$ (see black $\star$ symbols in Fig.~\ref{fig:charges}).
That is, we have power-law density correlations at finite $2k_F$ wavevectors---a manifestation of which
are the famous Friedel oscillations common in metals---even in a Mott insulator!

\def\dimersCaption
{
\textbf{Dimer structure factor: Period-2 valence bond solid order in the strong Mott insulator.}
Measurements of the dimer structure factor, $\langle \mathcal{B}_q\mathcal{B}_{-q}\rangle$,
show the emergence of a C0S0 period-2 valence bond solid for $U/t\gtrsim5.0$.
Its long-range order is very clearly demonstrated by the prominent Bragg peaks at $q=\pi$,
as shown in the inset.  Data correspond to the same $U/t$ values and color scheme as
in Figs.~\ref{fig:charges} and \ref{fig:spins}.
In the main panel (inset), we show data only for the metal and spinon metal (valence bond solid)
corresponding to values $U/t<5.0$ ($U/t\geq5.0$).
}

\begin{figure}[t]
\includegraphics[width=1.0\columnwidth]{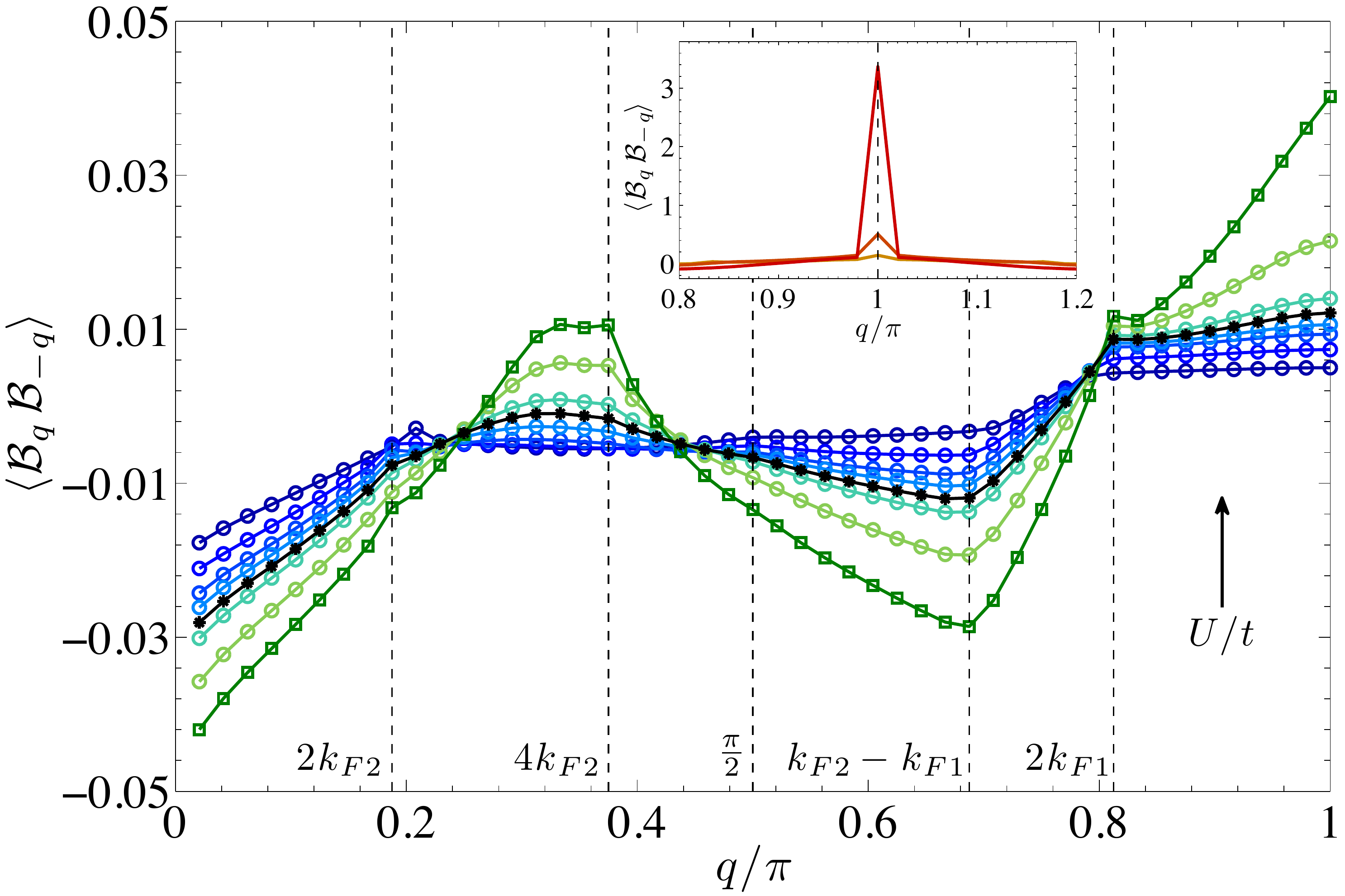} 
\caption{\dimersCaption}
\label{fig:dimers}
\end{figure}

Indeed, this remarkable result is expected in the two-band spinon metal theory,
where, as with the spin operator, the slowly varying part of the density operator at wavevectors
$Q=2k_{F1}, 2k_{F2}, \pi/2$ again contains $\theta_{\rho+}$
(but not the wildly fluctuating conjugate field $\varphi_{\rho+}$), i.e., $\delta n_Q\sim e^{\pm i\theta_{\rho+}}(\cdots)$.
Thus, we should even expect the scaling dimension of the density operator at these wavevectors
to be \emph{reduced} due to the same Amperean attraction mechanism responsible for enhancement of spin correlations
in Fig.~\ref{fig:spins}.  However, there are overriding nonuniversal amplitudes
that are expected to be small in a Mott insulator thus preventing observation
of this enhancement---this is likely the case in our data.
Furthermore, we see development of a feature, though apparently weak or with very small amplitude,
as anticipated, at a wavevector $q=4k_{F2}=-4k_{F1}$ (see black hexagram symbols in Fig.~\ref{fig:charges}).
This feature is again expected from theory and is actually a \emph{four-fermion}
contribution to the density operator~\cite{Sheng:2009up} (and thus is extremely weak at weak coupling).
Interestingly, all these power-law density correlations in our electronic two-band spinon metal
are a direct two-leg analog~\cite{Motrunich09_PRB_79_235120} of the charge Friedel oscillations
expected on the insulating side of the continuous Mott transition in higher dimensions,
as recently stressed by Mross and Senthil~\cite{Mross11_PRB_84_041102}.

Returning to the spin sector, we can use the small $q$ behavior of $\langle\mathbf{S}_{q}\cdot\mathbf{S}_{-q}\rangle$
to assess whether or not the spin sector is gapless in the realized phases.
In analogy with Eq.~(\ref{eq:dndnqSlope}), for a spin gapless state we have
\begin{equation}
\langle\mathbf{S}_{q}\cdot\mathbf{S}_{-q}\rangle = 3g_{\sigma+} |q|/2\pi~~\mathrm{as}~~q\rightarrow0,
\label{eq:sdotsqSlope}
\end{equation}
where $g_{\sigma+}$ is the ``Luttinger parameter'' associated with the overall spin mode $\theta_{\sigma+}$,
which for a gapless SU(2) invariant fixed point is necessarily unity:  $g_{\sigma+}=1$
(see Appendix~\ref{sec:observables_theory} and also, e.g., Refs.~\onlinecite{Giamarchi, Sedlmayr13_PRB_88_195113}).
In the top inset of Fig.~\ref{fig:spins}, we show $\langle\mathbf{S}_{q}\cdot\mathbf{S}_{-q}\rangle/(3|q|/2\pi)$,
where we see that for free electrons $g_{\sigma+}=1$,
while increasing $U/t$ pushes the $L=96$ estimate of $g_{\sigma+}$ above unity---this
increasing trend continues until $U/t\simeq4.0$, i.e., well beyond the Mott critical value of $U/t=1.6$.
This robust increasing measurement of $g_{\sigma+}>1$
(we expect $g_{\sigma+}\rightarrow1$ as $L\rightarrow\infty$)
well into the insulator is a strong indicator that the spin is gapless
on both the metallic and insulating sides of the Mott transition,
lending strong credence that we are indeed observing the sought-after
C2S2$\rightarrow$C1S2 scenario described above.
In Appendix~\ref{sec:bosonization}, we discuss these results in more depth and make comparisons to how
$g_{\sigma+}$ behaves in the on-site $t$-$t'$-$U$ Hubbard model at $\kappa=0$.

\def\electronCaption
{
\textbf{Electronic momentum distribution function: Disappearance of the Fermi surface.}
A dense scan of the electron momentum distribution function, $\langle
c_{q\alpha}^\dagger c_{q\alpha}\rangle$, over $U/t$ shows the gradual
disappearance of the Fermi surface with increasing interactions, as we move from
a two-band C2S2 metal ($U/t<1.6$) across the insulating C1S2 spinon metal (SM) ($1.6<U/t\lesssim5.0$)
to the C0S0 valence bond solid insulator ($U/t\gtrsim5.0$).
Vertical dashed lines mark the Fermi points (see Fig.~\ref{fig:dispersion}),
and the data is for the same $L=96$ site system as shown in Figs.~\ref{fig:charges}-\ref{fig:dimers}.
}

\begin{figure}[t]
\vspace{-0.12in}
\includegraphics[width=1.0\columnwidth]{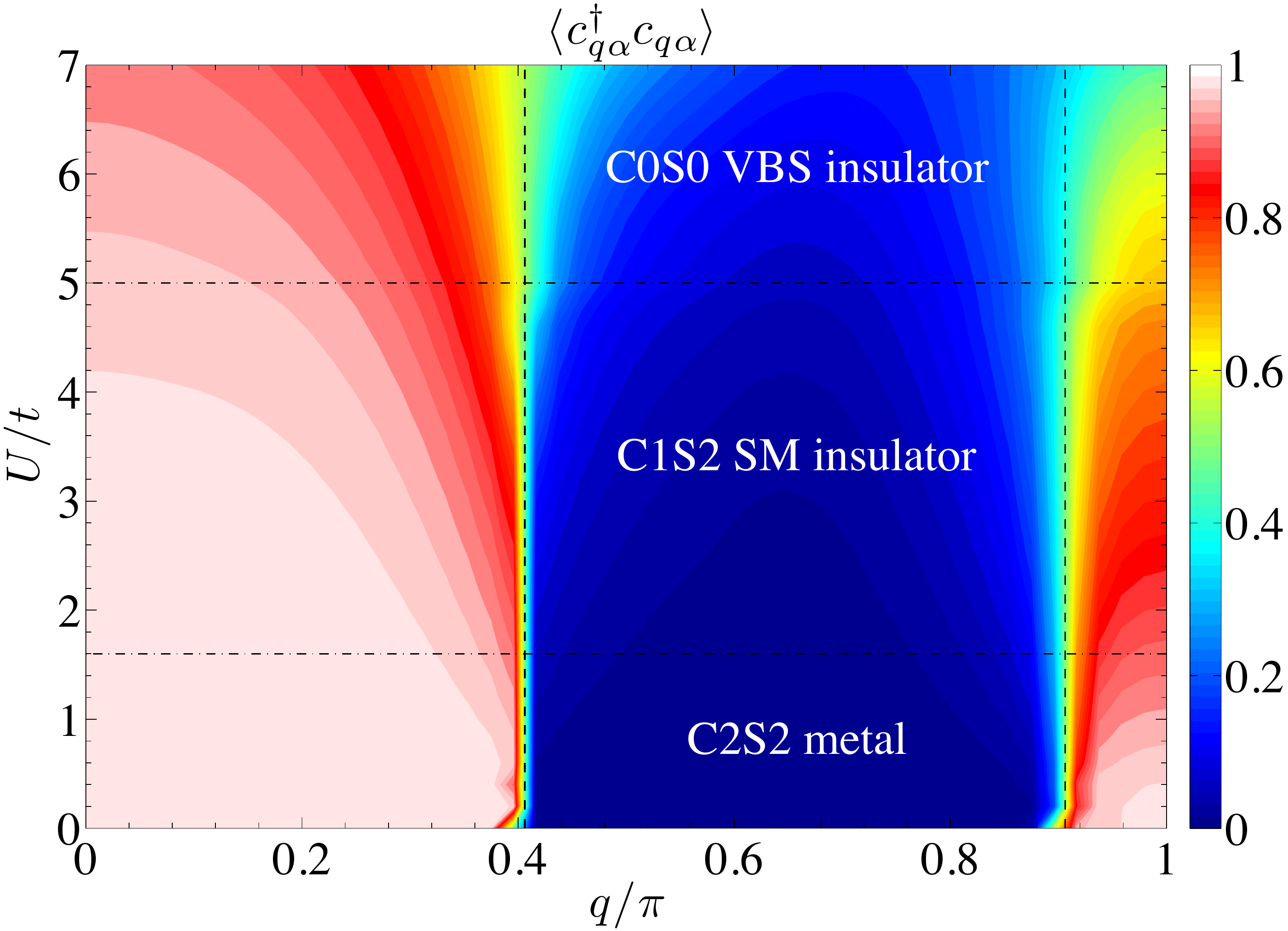}
\caption{\electronCaption}
\label{fig:electron}
\end{figure}

Eventually, above $U/t\simeq5.0$ we see that $g_{\sigma+}$ drops below unity and
$\langle\mathbf{S}_{q}\cdot\mathbf{S}_{-q}\rangle\sim q^2$ for small $q$,
indicating the opening of a spin gap.
We identify this strong Mott insulating phase as a fully gapped (C0S0) period-2 valence bond solid,
which is continuously connected to the dimerized phase realized by the $J_1$-$J_2$
Heisenberg model~\cite{White96_PRB_54_9862}
(and also the on-site $t$-$t'$-$U$ Hubbard model at large $U/t$~\cite{Tocchio10_PRB_81_205109}).
To this end, we turn to the dimer structure factor in Fig.~\ref{fig:dimers}.
In the inset, we indeed see clear Bragg peaks developing in $\langle \mathcal{B}_q\mathcal{B}_{-q}\rangle$
at $q=\pi$ for $U/t\gtrsim5.0$, hence strongly indicative of period-2 valence bond solid order.
Furthermore, the operator content of the density, $\delta n(x)$, and bond energy, $\mathcal{B}(x)$,
are identical at all wavevectors except $\pi$
(see Ref.~\onlinecite{Sheng:2009up} and Appendix~\ref{sec:observables_theory}).
Thus, in the gapless phases (C2S2 and C1S2) we expect singularities
in $\langle \mathcal{B}_q\mathcal{B}_{-q}\rangle$ at the same ``$2k_F$'' wavevectors
for which we find singularities in $\langle\delta n_{q}\delta n_{-q}\rangle$ (see Fig.~\ref{fig:charges}).
Indeed, in the main plot of Fig.~\ref{fig:dimers} we clearly see features
in $\langle \mathcal{B}_q\mathcal{B}_{-q}\rangle$ at $q=2k_{F1}, 2k_{F2}, k_{F2}-k_{F1}$, and $4k_{F2}$. 
Once in the putative C1S2 insulator, these features are more apparent in
$\langle \mathcal{B}_q\mathcal{B}_{-q}\rangle$ than in $\langle\delta n_{q}\delta n_{-q}\rangle$
since the latter are expected to have small amplitudes in a Mott insulator.
In our data, this is especially true at wavevectors $2k_{F2}$ and $4k_{F2}$,
the latter of which is the very nontrivial four-fermion contribution discussed above.

Finally, we discuss the behavior of the electron momentum distribution function
$\langle c_{q\alpha}^\dagger c_{q\alpha}\rangle$ as shown for a dense scan of $U/t$
values in Fig.~\ref{fig:electron}.  Beyond the Mott transition, when the field $\theta_{\rho+}$
gets pinned, we expect the electron Green's function to decay exponentially
so that the power-law singularities in $\langle c_{q\alpha}^\dagger c_{q\alpha}\rangle$
at the four Fermi points $q=\pm k_{F1}, \pm k_{F2}$ become gapped.
While it is not exceedingly apparent that finite correlation lengths emerge at the Fermi points
when we cross the Mott transition at $U/t=1.6$
(as determined from $g_{\rho+}$ measurements---see Fig.~\ref{fig:charges}),
we believe this is another manifestation of the large charge correlation lengths
present in the exotic C1S2 insulator.  Deep into the putative C1S2 phase though,
e.g., for $U/t\simeq4.0$, finite correlation lengths are more apparent.

\section{\uppercase{Discussion and outlook}}

In this paper, we have explored the Mott transition between a metal and a quantum spin liquid,
presenting strong evidence through large-scale DMRG simulations in quasi-1D
that such a continuous transition can be realized in reasonable electronic models.
Our study is strongly motivated by recent experiments on the
quasi-two-dimensional organic materials \kappaET~and~\dmit, each of which is a quantum spin liquid
that can be driven through a Mott transition to a Fermi liquid under pressure.
We believe our simulations of an extended Hubbard model---a model well-motivated by
recent \emph{ab initio} calculations~\cite{Imada09_JPSJ_78_083710, Hotta14_PRB_89_045102} on \kappaET---represent
an important first step toward numerically characterizing this transition.
While our study is restricted to the two-leg triangular strip,
it does show the universal physics of a clear and direct quasi-1D analog of the
continuous Mott metal-to-spin liquid transition in two dimensions~\cite{Senthil:2008ki}.
It is important to point out that the physics realized above is markedly distinct from the
well-known strictly one-dimensional case where a single nested pair of Fermi points gaps out
at infinitesimal $U/t=0^+$.  In our case, we have \emph{two unnested} pairs of Fermi points which gap out
simultaneously at some finite and intermediate value of $U/t$.
Thus, qualitatively speaking, our results are remarkably reminiscent of what would happen
in full two dimensions where the entire Fermi surface gaps out at the transition~\cite{Senthil:2008ki}.

Just as importantly, our calculations also elucidate the remarkable properties of the spin-liquid state
stabilized on the insulating side.  In many ways, this electronic ``spinon metal'' weak Mott insulator,
as realized in our model, behaves very much like a metal on length scales shorter than the charge correlation length,
and indeed exhibits long-distance density and spin correlations reminiscent of the nearby metallic phase
(see Figs.~\ref{fig:charges} and \ref{fig:spins}).  It is precisely this striking similarity between
the metallic and insulating states---in basically all properties except the finite charge correlation length
in the latter---which makes a continuous Mott metal-insulator transition plausible, perhaps even likely.

Going forward, it would clearly be desirable to move towards two dimensions and explore
the Mott transition in models such as Eq.~(\ref{eq:hamiltonian}) on wider ladders and eventually in full 2D,
with the goal to make real connections with the actual experiments~\cite{Kurosaki:2005ji, Kato06_DMITmott, Kanoda12_fragnetsConfTalk, Kanoda15_NaturePhys_11_221}.
In the end, the transition may turn out to not be continuous but instead be weakly first order,
as is perhaps realized in \kappaET.
Still, our numerical calculations presented here,
as well as the recent field theoretic work of Senthil \emph{et al.},
suggest that a continuous Mott transition in the ($d+1$)D XY universality class
between a metal and quantum spin liquid is a very real, exciting possibility.

\def\acknowledgementsText
{
We would like to thank Adrian del Maestro, Arun Paramekanti, Federico Becca,
Hsin-Hua Lai, David Mross, William Witczak-Krempa, T. Senthil, and K. Kanoda for useful discussions.
R.V.M. is especially grateful to Max Metlitski for help on the RG calculation discussed
in Appendix~\ref{sec:RG}. %Sec.~\ref{sec:RG}.
This work was supported by the NSF under grants DMR-1101912 (R.V.M. and M.P.A.F.),
PHY11-25915 (R.G.M.), and DMR-1206096 (O.I.M.); MICINN through grant FIS2009--13520 (I.G.).
R.G.M acknowledges support from NSERC, the Canada Research Chair program, the John Templeton Foundation, and the Perimeter Institute (PI) for Theoretical Physics.  Research at PI is supported by the Government of Canada through Industry Canada and by the Province of Ontario through the Ministry of Economic Development \& Innovation.
We also acknowledge support by the Caltech Institute of Quantum Information and Matter,
an NSF Physics Frontiers Center with the support of the Gordon and Betty Moore Foundation (O.I.M. and M.P.A.F.).
This work was made possible by the computing facilities of the Center for Scientific Computing from the CNSI,
MRL: an NSF MRSEC award (DMR-1121053), and an NSF grant (CNS-0960316); and CESGA.
I.G. acknowledges hospitality from the University of California, Santa Barbara,
during a research stay when much of this work was done.
}

\acknowledgements \acknowledgementsText

\appendix

\section{Details of DMRG calculations and observables} \label{sec:observables_def}

We use large-scale DMRG calculations to calculate ground state properties of our
model Hamiltonian, Eqs.~(\ref{eq:hamiltonian})-(\ref{eq:modelV}), on finite-size chains of length $L$ sites.
While we have performed simulations with both open and periodic boundary conditions, %(OBC and PBC)
we find the latter to be preferable for our model in spite of the well-known
more challenging convergence properties with periodic boundaries in DMRG calculations.
The long-ranged nature of our interaction potential [Eq.~(\ref{eq:modelV})], however,
makes open boundaries problematic.  The issue is that all interactions up to fourth neighbor
are chosen to scale with the overall Hubbard strength $U$,
so that, at least for the parameters chosen in our study, it is energetically favorable
for the end sites of an open chain to become doubly occupied at large $U/t$.  That is,
even though the system then has to pay very large on-site $U$ on the end sites,
it gains significant energy by not having to pay as substantial $V_1$ to $V_4$.
Therefore, for the calculations on the extended Hubbard model presented in the main text,
we have employed periodic boundary conditions.

To numerically characterize the ground state properties of the system with the DMRG,
we calculate the density structure factor $\langle \delta n_q \delta n_{-q} \rangle$,
the spin structure factor $\langle \mathbf{S}_q\cdot\mathbf{S}_{-q}\rangle$,
the dimer structure factor $\langle\mathcal{B}_q\mathcal{B}_{-q} \rangle$,
and the electron momentum distribution function $\langle c^\dagger_{q\alpha}c_{q\alpha}\rangle$
(where $\alpha=\,\uparrow, \downarrow$ with no implied summation).
In each case, the structure factor is defined as the Fourier transform of the associated two-point function.
Specifically, we have
\begin{equation}
\langle \delta n_q \delta n_{-q} \rangle = \frac{1}{L} \sum_{x,x'} e^{-iq(x-x')}
\langle \delta n(x) \delta n(x') \rangle,
\label{eq:dndnq}
\end{equation}
\begin{equation}
\langle \mathbf{S}_q\cdot\mathbf{S}_{-q}\rangle = \frac{1}{L} \sum_{x,x'} e^{-iq(x-x')}
\langle\mathbf{S}(x)\cdot\mathbf{S}(x')\rangle,
\end{equation}
\begin{equation}
\langle \mathcal{B}_q\mathcal{B}_{-q}\rangle = \frac{1}{L} \sum_{x,x'} e^{-iq(x-x')}
\langle \mathcal{B}(x)\mathcal{B}(x') \rangle,
\label{eq:BBq}
\end{equation}
\begin{equation}
\langle c^\dagger_{q\alpha}c_{q\alpha}\rangle = \frac{1}{L} \sum_{x,x'} e^{-iq(x-x')}
\langle c^\dagger_\alpha(x)c_\alpha(x')\rangle,
\end{equation}
where $n(x)\equiv\sum_{\alpha=\up,\dn} c^{\dagger}_{\alpha}(x)c_{\alpha}(x)$ is the number operator
[with $\delta n(x)\equiv n(x)-\langle n(x)\rangle$],
$\mathbf{S}(x)\equiv\frac{1}{2}\sum_{\alpha,\beta}c^{\dagger}_{\alpha}(x)\bm{\sigma}_{\alpha\beta}c_{\beta}(x)$ is the spin operator,
and $\mathcal{B}(x)\equiv\mathbf{S}(x)\cdot\mathbf{S}(x+1)$ is the bond energy operator.
For simplicity, we set $\langle \mathcal{B}(x)\mathcal{B}(x') \rangle=0$ if $\mathcal{B}(x)$ and $\mathcal{B}(x')$
share common sites~\cite{Sheng:2009up}.
When presenting all structure factor measurements, we only show data for $q\geq0$
since the measurements are symmetric about $q=0$.

For the dimer structure factor in Eq.~(\ref{eq:BBq}), we do not subtract a product of local averages from the
$\langle \mathcal{B}(x)\mathcal{B}(x') \rangle$ correlations as we do, e.g., for the density structure factor
in Eq.~(\ref{eq:dndnq}). The main reason for this choice is that at large $U/t\gtrsim5.0$ our DMRG calculations,
even with periodic boundary conditions, have a tendency to get ``stuck'' in one of the two possible
symmetry broken period-2 VBS patterns, giving a rather strong period-2 signal in the local expectation value
$\langle\mathcal{B}(x)\rangle=\langle\mathbf{S}(x)\cdot\mathbf{S}(x+1)\rangle$.
This is likely due to the somewhat awkward way in which periodic boundaries
are implemented in a traditional DMRG setup which treats the end sites on a different footing.
Fourier transforming
$\langle \mathcal{B}(x)\mathcal{B}(x') \rangle - \langle \mathcal{B}(x)\rangle\langle\mathcal{B}(x') \rangle$
then washes out the Bragg peaks preasent at $q=\pi$.
Hence, we just use $\langle \mathcal{B}(x)\mathcal{B}(x') \rangle$
as the real-space two-point function and exclude plotting $\langle \mathcal{B}_q\mathcal{B}_{-q}\rangle$ at $q=0$.
This both well captures the obvious Bragg peaks at $q=\pi$ in the C0S0 and also gives very clear power-law
singularities at the various ``$2k_F$'' wavevectors as expected in the C1S2 insulator
(see Fig.~\ref{fig:dimers}, Appendix~\ref{sec:observables_theory}, and Ref.~\onlinecite{Sheng:2009up}).

More generally, we find that the averaging done in our Fourier transforms when summing over
both $x$ and $x'$ in Eqs.~(\ref{eq:dndnq})-(\ref{eq:BBq}) does an effective job of representing the
structure factors in cases where, due to slight lack of convergence in the DMRG ground state,
the two-point functions depend on both the separation distance $x-x'$ and the ``origin'' $x'$.
(Of course, for a perfectly translationally invariant state the two-point functions depend only on $x-x'$.)

In our DMRG calculations, we keep up to $m=6000$ states and perform at least 6 finite-size sweeps
which results in a density matrix truncation error of on the order of $10^{-5}$ or smaller.
All measurements are well-converged to the extent necessary to establish the statements made in the main text.
To get a feel for the difficulty encountered in obtaining highly accurate data on the stiffness parameters
$g_{\rho+}$ and $g_{\sigma+}$ (see the main text and Appendix~\ref{sec:bosonization} below),
one can observe the data in the insets of Figs. \ref{fig:charges} and \ref{fig:spins}
at the free electron point $U/t=0$---basically the most challenging point for the DMRG.
For free electrons, we should have $g_{\rho+} = g_{\sigma+} = 1$.
We see that there is a rather severe error at the first allowed momentum $q=2\pi/L$,
yet the error for momenta $q > 2\pi/L$ is very acceptable, on the order of $1\%$ or less.

\section{Luttinger liquid description and solution by bosonization} \label{sec:bosonization}

In this section, we spell out the effective low-energy description of the C2S2 metal
and C1S2 spinon metal and intervening Kosterlitz-Thouless (KT)-like Mott transition,
focusing on those aspects of the theory most relevant to the DMRG results presented in the main text.
Some aspects of our presentation follow that of Refs.~\onlinecite{Sheng:2009up, Motrunich10_PRB_81_045105}.

\subsection{Long-wavelength description of C2S2 metal and C1S2 spinon metal} \label{sec:longwave}

Consider noninteracting electrons at half-filling on the two-leg triangular strip (see Fig.~\ref{fig:ladder}).
When viewed as a 1D chain with first-neighbor and second-neighbor hopping,
$t$ and $t'$, the electron dispersion is given by (see also Fig.~\ref{fig:dispersion})
\begin{equation}
\epsilon(q) = -2t\cos(q) - 2t'\cos(2q) - \mu.
\end{equation}
For $t'/t>0.5$, which is the case of interest here, the ground state
consists of two disconnected Fermi seas (bands) which we label by $a=1,2$.
We take the convention that the Fermi velocities $v_{Fa}$ are positive (negative)
for electrons moving near $k_{Fa}$ ($-k_{Fa}$), corresponding to right and left movers, respectively.
Furthermore, taking the system to be at half-filling gives the sum rule $k_{F1}+k_{F2}=-\pi/2\mod 2\pi$.

As usual~\cite{Giamarchi}, we take the low-energy continuum limit and expand
the electron operator in terms of slowly varying continuum fields at the four Fermi points:
\begin{equation}
c_\alpha(x) = \sum_{a,P} e^{iPk_{Fa}x}c_{Pa\alpha}\,,
\label{eq:contfields}
\end{equation}
where $\alpha=\,\uparrow,\downarrow$ denotes the electron spin,
and the sum runs over $a=1,2$ for the two Fermi seas
and $P=R/L=+/-$ for the right and left moving electrons at the Fermi points of each Fermi sea.
Although not written explicitly, the continuum fields of course depend on position $x$:
$c_{Pa\alpha}=c_{Pa\alpha}(x)$.

Next, we bosonize~\cite{Giamarchi} the continuum fields according to
\begin{equation}
c_{Pa\alpha} = \eta_{a\alpha} e^{i(\varphi_{a\alpha} + P\theta_{a\alpha})},
\label{eq:bosonization}
\end{equation}
where $\varphi_{a\alpha}$ and $\theta_{a\alpha}$ are the canonically conjugate bosonic
phase and phonon fields, respectively.  Specifically, we have
\begin{align}
&[\varphi_{a\alpha}(x), \varphi_{b\beta}(x')] = [\theta_{a\alpha}(x), \theta_{b\beta}(x')] = 0, \label{eq:commute} \\
&[\varphi_{a\alpha}(x), \theta_{b\beta}(x')] = i\pi\delta_{ab}\delta_{\alpha\beta}\Theta(x-x'). \label{eq:densphase}
\end{align}
The fields $\eta_{a\alpha}$ are the Klein factors, i.e., Majorana fermions
$\{\eta_{a\alpha},\eta_{b\beta}\}=2\delta_{ab}\delta_{\alpha\beta}$, which
are necessary to ensure the correct anticommutation relations
among different fermionic species $a\alpha$.
Finally, the slowly varying component of the electron density
is given by the derivative of the $\theta_{a\alpha}$ fields:
$\rho_{a\alpha} = \sum_{P=\pm} c^\dagger_{Pa\alpha}c_{Pa\alpha} = \partial_x\theta_{a\alpha}/\pi$,
where $c^\dagger_{Pa\alpha}c_{Pa\alpha} = \partial_x(\theta_{a\alpha} + P\varphi_{a\alpha})/(2\pi)$.
Hence, Eq.~(\ref{eq:densphase}) is essentially a statement of the density-phase uncertainty relation:
$[\rho(x),\varphi(x')]=i\delta(x-x')$.

Next, we linearize about the Fermi points and express the problem in terms of the bosonized fields introduced above.
Working in the Euclidean path integral formalism,
the low-energy continuum Lagrangian density for the two-band noninteracting electron gas then reads:
\begin{equation}
\Lag_\mathrm{free} = \Ham_\mathrm{free}
+  \sum_{a,\alpha}\frac{i}{\pi}(\partial_x\theta_{a\alpha})(\partial_\tau\varphi_{a\alpha}),
\end{equation}
where
\begin{equation}
\Ham_\mathrm{free} =  \sum_{a,\alpha}\frac{v_{Fa}}{2\pi}\left[(\partial_x\theta_{a\alpha})^2 + (\partial_x\varphi_{a\alpha})^2\right].
\end{equation}

We now introduce the ``charge'' and ``spin'' modes for each band:
\begin{equation}
\theta_{a\rho/\sigma} \equiv \frac{1}{\sqrt{2}}\left( \theta_{a\uparrow} \pm \theta_{a\downarrow} \right),
\label{eq:spinchargefields}
\end{equation}
and the ``overall'' and ``relative'' combinations with respect to the two bands:
\begin{equation}
\theta_{\mu\pm} \equiv \frac{1}{\sqrt{2}}\left( \theta_{1\mu} \pm \theta_{2\mu} \right),
\label{eq:pmfields}
\end{equation}
where $\mu=\rho,\sigma$.
Fields analogous to Eqs.~(\ref{eq:spinchargefields}) and (\ref{eq:pmfields}) are also defined for the $\varphi$'s.
These newly defined fields satisfy the same canonical commutation relations as the original fields
[Eqs.~(\ref{eq:commute})-(\ref{eq:densphase})].
The free-electron Lagrangian $\Lag_\mathrm{free}$ then as usual decouples into charge and spin sectors:
\begin{equation}
\Lag_\mathrm{free}=\Lag_\mathrm{free}^\rho + \Lag_\mathrm{free}^\sigma,
\label{eq:freefermionLag}
\end{equation}
where
\begin{align}
\Lag_\mathrm{free}^\mu &= \Ham_\mathrm{free}^\mu + \sum_{a}\frac{i}{\pi}(\partial_x\theta_{a\mu})(\partial_\tau\varphi_{a\mu}), \\
\Ham_\mathrm{free}^\mu &= \sum_{a}\frac{v_{Fa}}{2\pi}\left[(\partial_x\theta_{a\mu})^2 + (\partial_x\varphi_{a\mu})^2\right].
\end{align}

We are finally in position to discuss interactions.  In the interacting C2S2 Luttinger liquid,
the fixed-point theory is similar to Eq.~(\ref{eq:freefermionLag}), i.e.,
\begin{equation}
\Lag_\mathrm{C2S2} = \Lag_\mathrm{C2S2}^\rho + \Lag_\mathrm{C2S2}^\sigma,
\label{eq:C2S2Lag}
\end{equation}
except we have general mode velocities and, in the charge sector, nontrivial Luttinger parameters.
For convenience in the discussion that follows, in the charge sector
we work in the $\rho\pm$ basis of Eq.~(\ref{eq:pmfields})
and write the most general charge sector Lagrangian as
\begin{align}
\Lag_\mathrm{C2S2}^\rho &= \Ham_\mathrm{C2S2}^\rho + \frac{i}{\pi}\partial_x\mathbf{\Theta}^T\cdot\partial_\tau\mathbf{\Phi},
\label{eq:LrhoFull} \\
\Ham_\mathrm{C2S2}^\rho &= \frac{1}{2\pi}\left[\partial_x\mathbf{\Theta}^T\cdot \mathbf{A}\cdot\partial_x\mathbf{\Theta} + 
\partial_x\mathbf{\Phi}^T\cdot\mathbf{B}\cdot\partial_x\mathbf{\Phi}\right], \label{eq:LrhoFullHami}
\end{align}
where $\mathbf{\Theta}^T\equiv(\theta_{\rho+},\theta_{\rho-})$ and $\mathbf{\Phi}^T\equiv(\varphi_{\rho+},\varphi_{\rho-})$;
$\mathbf{A}$ and $\mathbf{B}$ are symmetric, positive definite 2x2 matrices which encode interactions.
Note that even for free electrons, if $v_{F1}\neq v_{F2}$, the charge sector is
not diagonal in the $\rho\pm$ basis, i.e.,
$A_{12}=A_{21}\neq0$, $B_{12}=B_{21}\neq0$,
and in general the interacting C2S2 metal will have coupled
$\rho+$ and $\rho-$ modes~\cite{Motrunich10_PRB_81_045105}.

For the spin sector, we stay in the band basis $a=1,2$ and write
\begin{align}
\Lag_\mathrm{C2S2}^\sigma &= \Ham_\mathrm{C2S2}^\sigma + \sum_{a}\frac{i}{\pi}(\partial_x\theta_{a\sigma})(\partial_\tau\varphi_{a\sigma}), \\
\Ham_\mathrm{C2S2}^\sigma &= \sum_{a}\frac{v_{a\sigma}}{2\pi}\left[\frac{1}{g_{a\sigma}}(\partial_x\theta_{a\sigma})^2 + g_{a\sigma}(\partial_x\varphi_{a\sigma})^2\right]. \label{eq:spinHami}
\end{align}
SU(2) invariance dictates only trivial Luttinger parameters in the spin sector, i.e.,
$g_{1\sigma}=g_{2\sigma}=1$ (see Appendix~\ref{sec:observables_theory}),
but we keep them general in Eq.~(\ref{eq:spinHami}) for further analysis below.
Our representation of the spin sector here is somewhat schematic in that allowed strictly marginal chiral interactions
will couple the bare spin modes [Eq.~(\ref{eq:spinchargefields})] in the quadratic part of the C2S2 action.
However, the resulting $\Ham_\mathrm{C2S2}^\sigma$ is symmetric
under interchanging $\theta_{a\sigma}\leftrightarrow\varphi_{a\sigma}$
and so can easily be brought back to diagonal form via a simple orthogonal transformation
which acts identically on the $\theta_{a\sigma}$ and $\varphi_{a\sigma}$ fields,
hence keeping the Luttinger parameters at their trivial values.
Thus, for the quadratic part of the C2S2 fixed-point theory,
Eq.~(\ref{eq:spinHami}) is completely general for our purposes.
Interestingly, the full C2S2 fixed-point theory also contains a strictly marginal chiral interband scattering term
of the form $(\Ham_\mathrm{chiral}^\sigma)_\perp \sim \cos(2\varphi_{\sigma-})\cos(2\theta_{\sigma-})$,
which is nonharmonic~\cite{Sedlmayr13_PRB_88_195113}.
However, we expect that the presence of this, presumably exactly marginal, nonharmonic chiral interaction will not
quantitatively alter the spin sector at the C2S2 (and C1S2; see below) fixed point---at
least with respect to the Luttinger parameters and contributions to the scaling dimensions of various operators
(see Appendix~\ref{sec:observables_theory}).
In fact, assuming that $(\Ham_\mathrm{chiral}^\sigma)_\perp$ is exactly marginal already \emph{implies}
trivial spin sector Luttinger parameters, $g_{1\sigma}=g_{2\sigma}=1$, which is encouraging.

In addition to such strictly marginal interactions, there are many nonchiral interactions allowed
by symmetry which may be added to Eq.~(\ref{eq:C2S2Lag})
and potentially destabilize the C2S2 theory described above.
To connect to a given microscopic Hamiltonian,
a common approach is to employ a weak-coupling renormalization group (RG) scheme.
That is, one can project the microscopic interactions
onto all continuum symmetry-allowed interactions and read off initial conditions for all such couplings;
these initial conditions can then be subsequently used in a
controlled RG analysis valid for weak microscopic coupling ${U/t\ll 1}$.
Then, bosonizing the four-fermion interactions---particularly
those that may flow to strong coupling, hence destabilizing the ``mother'' C2S2---emits
a direct physical interpretation of the resulting phase.
This is the approach pioneered many years ago in Ref.~\onlinecite{Balents96_PRB_53_12133},
where it was shown (see also Ref.~\onlinecite{Gros01_PRB_64_113106})
that for the on-site $t$-$t'$-$U$ Hubbard model,
the C2S2 metal is generally unstable at weak repulsive interactions to the opening of a spin gap.
The basic idea is that the RG flow equations---which are indeed rather complicated for the two-band system
and in general require a detailed numerical analysis---have a tendency to eventually drive \emph{attractive} divergent couplings
in the spin sector (e.g., the terms denoted $g_{a\sigma}$ in Ref.~\onlinecite{Balents96_PRB_53_12133}
or, equivalently, $\lambda_{aa}^\sigma$ in Ref.~\onlinecite{Motrunich10_PRB_81_045105}).
These divergent couplings conspire to gap out all modes except the overall conducting charge mode $\theta_{\rho+}$,
leaving a one-mode C1S0 conducting Luttinger liquid, essentially the quasi-1D analog of a superconductor.

However, this spin-gap tendency is not unavoidable.
For example, one can fight such pairing tendencies by adding longer-ranged repulsion
to the model Hamiltonian.  This approach was recently explored systematically in
Ref.~\onlinecite{Motrunich10_PRB_81_045105}, where it was shown that the
C2S2 metal occupies a substantial portion of the weak-coupling phase diagram
for the model considered in our work:  Eqs.~(\ref{eq:hamiltonian})-(\ref{eq:modelV}).
Stability of the C2S2 metal at weak coupling indeed seems to be a necessary component
for realizing the C2S2$\rightarrow$C1S2 Mott transition presented numerically in the main text,
and we buttress off the weak-coupling phase diagram presented in Ref.~\onlinecite{Motrunich10_PRB_81_045105}
when selecting the specific parameters of our model Hamiltonian.

Finally, as stressed in the main text, our Mott transition is driven at strong interactions by an
\emph{eight-fermion} umklapp term wherein both spin-up and spin-down electrons
are scattered across each Fermi sea (see Fig.~\ref{fig:dispersion}):
\begin{equation}
\mathcal{H}_8 = u(c^\dagger_{R1\uparrow}c^\dagger_{R1\downarrow}c^\dagger_{R2\uparrow}c^\dagger_{R2\downarrow}
c_{L1\uparrow}c_{L1\downarrow}c_{L2\uparrow}c_{L2\downarrow} + \mathrm{H.c.}),
\label{eq:H8cs}
\end{equation}
which when written in terms of the bosonized fields simply becomes a cosine of the overall charge field $\theta_{\rho+}$:
\begin{equation}
\mathcal{H}_8 = 2u\cos(4\theta_{\rho+}).
\label{eq:H8}
\end{equation}
The C1S2 spinon metal spin liquid corresponds to relevance of $\mathcal{H}_8$ so that $u$ flows
to strong coupling.  That is, the field content of the C1S2 fixed-point theory looks identical to that of C2S2
but with a massive overall charge mode $\theta_{\rho+}$.  Specifically, we have
\begin{equation}
\Lag_\mathrm{C1S2} = \Lag_\mathrm{C1S2}^\rho + \Lag_\mathrm{C1S2}^\sigma,
\end{equation}
where the ``charge sector'' now only contains the $\rho-$ mode:
\begin{align}
\Lag_\mathrm{C1S2}^\rho &= \Ham_\mathrm{C1S2}^\rho + \frac{i}{\pi}\partial_x\theta_{\rho-}\partial_\tau\varphi_{\rho-},
\label{eq:LrhoC1S2} \\
\Ham_\mathrm{C1S2}^\rho &= \frac{v_{\rho-}}{2\pi}\left[\frac{1}{g_{\rho-}}(\partial_x\theta_{\rho-})^2 + 
g_{\rho-}(\partial_\tau\varphi_{\rho-})^2\right],
\end{align}
which physically represents gapless local current fluctuations,
and the spin sector formally reads the same as before:
\begin{equation}
\Lag_\mathrm{C1S2}^\sigma = \Lag_\mathrm{C2S2}^\sigma,
\end{equation}
still with trivial Luttinger parameters, $g_{1\sigma}=g_{2\sigma}=1$.
For an extensive discussion of the C1S2 phase with respect to its features and stability,
we refer the reader to Ref.~\onlinecite{Sheng:2009up}.

\subsection{Renormalization group analysis of the C2S2$\rightarrow$C1S2 Mott transition} \label{sec:RG}

We now present the details of the critical theory describing our Mott transition.
The theory is KT-like with a complication arising because the field $\theta_{\rho+}$, which is being gapped out,
is coupled to the field $\theta_{\rho-}$ in the Gaussian fixed-point action for the C2S2
[see Eq.~(\ref{eq:LrhoFullHami})], and $\theta_{\rho-}$ is massless on both sides of the transition.

From the above considerations, the charge sector Lagrangian describing the transition
between the C2S2 metal and C1S2 spinon metal reads
\begin{equation}
\mathcal{L} = \mathcal{L}_0 + \mathcal{L}_\mathrm{cos},
\end{equation}
where
\begin{equation}
\mathcal{L}_0 = \frac{1}{2\pi}\left[\partial_x\mathbf{\Theta}^T\cdot \mathbf{C}\cdot\partial_x\mathbf{\Theta} + 
\partial_\tau\mathbf{\Theta}^T\cdot\mathbf{D}\cdot\partial_\tau\mathbf{\Theta}\right] \label{eq:L0rho}
\end{equation}
is just $\Lag_\mathrm{C2S2}^\rho$ from Eq.~(\ref{eq:LrhoFull}) with the $\varphi$'s integrated out,
$\mathbf{\Theta}^T \equiv (\theta_{\rho+},\theta_{\rho-})$, and
\begin{equation}
\mathcal{L}_\mathrm{cos} = 2u\cos(n\theta_{\rho+})
\end{equation}
with $n=4$ is our eight-fermion umklapp term.
It is convenient to diagonalize the quadratic part of the theory $\mathcal{L}_0$ %[Eq.~(\ref{eq:L0rho})]
in a fashion similar to that described in Ref.~\onlinecite{Motrunich10_PRB_81_045105},
thus obtaining for the full theory
\begin{align}
&\mathcal{L}_0 = \frac{1}{2\pi}\sum_{i=1,2}\left[\frac{1}{v_i}(\partial_\tau\theta_{i})^2 + {v_i}(\partial_x\theta_{i})^2\right], \\
&\mathcal{L}_\mathrm{cos} = 2u\cos(n_1\theta_{1} + n_2\theta_{2}),
\end{align}
where we have absorbed the nontrivial Luttinger parameters of the two normal modes,
$\theta_1$ and $\theta_2$, into the real coefficients $n_1$ and $n_2$ via a rescaling of the fields.
While $\theta_1$ and $\theta_2$ are specific linear combinations of $\theta_{\rho+}$ and $\theta_{\rho-}$,
e.g., $n\theta_{\rho+} = n(c_1\theta_1 + c_2\theta_2) = n_1\theta_1 + n_2\theta_2$,
we do not spell out the details here, but instead refer the reader to the Appendix of
Ref.~\onlinecite{Motrunich10_PRB_81_045105} for a similar calculation.
Ultimately, this linear combination, as well as the velocities and Luttinger parameters
of the normal modes in the diagonalized system, are rather complicated, but still analytic,
functions of the original parameters $\mathbf{C}$ and $\mathbf{D}$ of the coupled system.

We have performed a renormalization group (RG) analysis of the above two-mode system,
obtaining the following leading-order KT-like (see below) flow equations for all couplings:
\begin{align}
\frac{dC_{11}}{d\ell} &= \frac{\pi n^2}{\Lambda^4\,v_1}I\left(\frac{v_2}{v_1},\frac{n_2^2}{4}\right)u^2, \label{eq:betaC11} \\
\frac{dD_{11}}{d\ell} &= \frac{\pi n^2}{\Lambda^4\,v_1^3}\left(\frac{v_2}{v_1}\right)^{-2n_2^2/4}
I\left(\frac{v_1}{v_2},\frac{n_2^2}{4}\right)u^2, \label{eq:betaD11} \\
\frac{du}{d\ell} &= \left[2 - \left(\frac{n_1^2}{4}+\frac{n_2^2}{4}\right)\right]u \label{eq:betau},
\end{align}
where
\begin{equation}
I(\alpha,\beta) \equiv \int_0^{2\pi}d\theta\frac{\cos^2\theta}{(\cos^2\theta + \alpha^2\sin^2\theta)^\beta} \geq 0.
\end{equation}

As with ordinary KT, the coupling $u$ renormalizes according to the scaling dimension of the cosine
with respect to the quadratic action,
\begin{equation}
\Delta[\cos(n\theta_{\rho+})] = \Delta[\cos(n_1\theta_1 + n_2\theta_2)] = \frac{n_1^2}{4} + \frac{n_2^2}{4},
\end{equation}
and obtaining its beta function, Eq.~(\ref{eq:betau}), can proceed in a textbook Wilsonian fashion~\cite{Giamarchi}.
However, renormalizing the parameters in $\mathcal{L}_0$ is significantly more involved
and depends on the specific regularization scheme employed.
First, note that since $\mathcal{L}_\mathrm{cos}$ contains only the field $\theta_{\rho+}$,
it cannot possibly renormalize any terms containing $\theta_{\rho-}$ to any order in perturbation theory;
hence, the only nonzero beta functions are those for the couplings $C_{11}$ and $D_{11}$.
The respective beta functions, Eqs.~(\ref{eq:betaC11}) and (\ref{eq:betaD11}),
were obtained using a field-theoretic approach~\cite{Amit80_JPhysA_13_585}
in which we consider insertions into correlation functions of the form
$\langle\partial_x\theta_i(x)\partial_x\theta_j(y)\rangle$, where $x$ and $y$ are points in our (1+1)D space-time.
At $\mathcal{O}(u^2)$, one has to integrate over two 2D points from two $u$ insertions, say $z$ and $z'$.
Indeed, as $z-z'$ becomes small, the integral diverges logarithmically, and we cut it off
at a short-distance scale $\Lambda^{-1}$.  We then compute corrections to
$\langle\partial_x\theta_i(x)\partial_x\theta_j(y)\rangle$ from posited ``counterterms'' in $\mathcal{L}_0$
which are chosen to exactly cancel the aforementioned logarithmic divergence.
This allows us, after an altogether somewhat lengthy calculation, to arrive at the above RG flow equations
for $C_{11}$ and $D_{11}$.

The case of vanishing $\theta_{\rho+}$-$\theta_{\rho-}$ coupling in Eq.~(\ref{eq:L0rho}) corresponds
to the limit $n_2\rightarrow0$, so that $\theta_1\propto\theta_{\rho+}$ and $C_{11}$ and $D_{11}$
renormalize at the same rate (up to an overall scale of $v_1^2$).
This of course corresponds to ordinary Kosterlitz-Thouless RG wherein only one parameter in
$\mathcal{L}_0$ renormalizes:  $\frac{d(g^{-1})}{d\ell}\sim u^2$,
with $g$ the single-mode Luttinger parameter~\cite{Giamarchi}.

In the general case, the beta functions for $C_{11}$ and $D_{11}$ involve highly nonuniversal content,
and thus we have not attempted a detailed study of the flows.  Still, the transition is
KT-like in nature except that two parameters (as opposed to one) in $\Lag_0$ are renormalized
by the single cosine, and the transition occurs when the scaling dimension of the cosine equals the space-time dimension:
$\Delta[\cos(n\theta_{\rho+})] = \frac{n_1^2}{4} + \frac{n_2^2}{4}=2$,
where $n_1$ and $n_2$ are functions of the parameters $\mathbf{C}$ and $\mathbf{D}$.

We can formally argue for the KT-like nature as follows.  From the start, we focus only on the flowing parameters
$C_{11}$, $D_{11}$, and $u$.  Let us denote the (non-negative) factors multiplying $u^2$ in the beta functions
for $C_{11}$ and $D_{11}$ as $A(C_{11}, D_{11})$ and $B(C_{11}, D_{11})$, respectively,
and also denote the coefficient of $u$ in the beta function for $u$ as $\Gamma(C_{11}, D_{11})$.
We emphasize that $A$, $B$, and $\Gamma$ are functions of $C_{11}$ and $D_{11}$,
which, while perhaps complicated functions, are analytical and not special.
As we vary in the $(C_{11}, D_{11})$ plane, we generically expect to find a line
where $\Gamma = 0$ separating regions where a small $u$ perturbation is relevant or irrelevant.
Let us consider one point on this line, $(C_{11}^{(0)}, D_{11}^{(0)})$,
and study small deviations $(\delta C_{11}, \delta D_{11})$ from this point.
The RG equations are, to leading order,
\begin{align}
\frac{d\,\delta C_{11}}{d\ell} &= A^{(0)} u^2, \\
\frac{d\,\delta D_{11}}{d\ell} &= B^{(0)} u^2, \\
\frac{du}{d\ell} &= \left(\alpha^{(0)} \delta C_{11} + \beta^{(0)} \delta D_{11}\right) u,
\end{align}
where $A^{(0)}$ and $B^{(0)}$ are the $A$ and $B$ functions evaluated at $(C_{11}^{(0)}, D_{11}^{(0)})$,
while $\alpha^{(0)}$ and $\beta^{(0)}$ are derivatives
$\partial\Gamma/\partial C_{11}$ and $\partial\Gamma/\partial D_{11}$ evaluated at the same point.
Deviations satisfying $\alpha^{(0)} \delta C_{11} + \beta^{(0)} \delta D_{11} = 0$
correspond to moving along the $\Gamma = 0$ line, while generic deviations will cut across this line.
Formally, we can change variables to
$r = \alpha^{(0)} \delta C_{11} + \beta^{(0)} \delta D_{11}$,
$s = -\beta^{(0)} \delta C_{11} + \alpha^{(0)} \delta D_{11}$,
which flow as
\begin{align}
\frac{dr}{d\ell} &= \left(\alpha^{(0)} A^{(0)} + \beta^{(0)} B^{(0)}\right) u^2, \\
\frac{ds}{d\ell} &= \left(-\beta^{(0)} A^{(0)} + \alpha^{(0)} B^{(0)}\right) u^2, \\
\frac{du}{d\ell} &= r u.
\end{align}
Thus, the flow equations for the $r$ and $u$ variables have familiar KT-like form
and subsequent standard analysis can kick in.  On the other hand, the flow of the $s$ variable
is simply slaved to $u$ and does not affect the KT analysis.

In principle, one should be able to confirm the KT universality class from the
numerical DMRG data, for example, by performing Weber-Minnhagen~\cite{Weber88_PRB_37_5986}
style fits to finite-size estimates of the scaling dimension of the cosine in the metallic phase
(essentially the stiffness in the XY model context; see also Appendix~\ref{sec:observables_theory} below).
However, this requires highly accurate data on large system sizes in the scaling regime,
which is currently prohibitive for our multimode electronic system (see Appendix~\ref{sec:observables_def}).
Also, it is not unreasonable to expect that the presence of two renormalizing parameters in $\Lag_0$,
instead of one, might make the finite-size effects more severe.
In the end though, this is a rather nonuniversal matter which we do not pursue further analytically.

\subsection{Observables and stiffness parameters} \label{sec:observables_theory}

To characterize the system, we have focused on the density structure factor, the spin structure factor,
the dimer structure factor, and the electron momentum distribution function as presented in the main text
and as defined in Appendix~\ref{sec:observables_def}.
In this section, we lay out the details of the bosonization treatment which allows us to use these
measurements, both at finite and zero wavevectors, to probe the nature of the Luttinger liquid phases
realized by our model Hamiltonian.

\subsubsection{Establishing the result $\Delta[\Ham_8] = 4g_{\rho+}$}

As stressed in the main text, we can directly measure the scaling dimension
of the eight-fermion umklapp term [see Eqs.~(\ref{eq:H8main}) and (\ref{eq:H8})] responsible for driving our Mott transition
by measuring the slope of the density structure factor at $q=0$ momentum [see Eq.~(\ref{eq:dndnqSlope})].
We now spell out how these two quantities, $\Delta[\Ham_8]$ and $g_{\rho+}$, are formally related.

The former is defined through the corresponding two-point function:
\begin{equation}
\left\langle e^{i4\theta_{\rho+}(x)} e^{-i4\theta_{\rho+}(0)} \right\rangle \sim \frac{1}{|x|^{2\Delta[\Ham_8]}},
\label{eq:DeltaH8def}
\end{equation}
where, for simplicity, we work at equal (imaginary) time such that $x$ is a spatial coordinate only.
Assuming that the system is in the C2S2 phase so that the charge sector is described
by the quadratic Lagrangian $\Lag_\mathrm{C2S2}^\rho$ of Eq.~(\ref{eq:LrhoFull}),
we can use a standard identity~\cite{Giamarchi} and write
\begin{equation}
\left\langle e^{i4\theta_{\rho+}(x)} e^{-i4\theta_{\rho+}(0)} \right\rangle
= e^{-\frac{4^2}{2}\left\langle[\theta_{\rho+}(x) - \theta_{\rho+}(0)]^2\right\rangle}.
\label{eq:H8intermediate}
\end{equation}
Now, the slowly varying component of the total electron density (measured relative to the average density)
is given by $\delta n(x) = 2\partial_x\theta_{\rho+}/\pi$, so that the long-wavelength contribution
to the density-density correlation function in real space is given by
\begin{equation}
\langle\delta n(x)\delta n(0)\rangle = \frac{4}{\pi^2}\partial_x\partial_{x'}\langle\theta_{\rho+}(x)\theta_{\rho+}(x')\rangle|_{x'=0}
+\cdots.
\end{equation}
The right-hand side can be obtained from Eq.~(\ref{eq:H8intermediate}) via straightforward manipulations,
which after invoking Eq.~(\ref{eq:DeltaH8def}) gives
\begin{align}
\langle\delta n(x)\delta n(0)\rangle = -\frac{\Delta[\Ham_8]}{2\pi^2}\frac{1}{x^2} + \cdots.
\label{eq:dndnxH8}
\end{align}

On the other hand, we define the slope of the momentum-space density structure factor
as $q\rightarrow0$ according to Eq.~(\ref{eq:dndnqSlope}), i.e.,
\begin{equation}
\langle\delta n_{q}\delta n_{-q}\rangle = \frac{2g_{\rho+}}{\pi}|q|,
\label{eq:dndnqSlope2}
\end{equation}
such that $g_{\rho+}=1$ corresponds to a two-band noninteracting electron gas. 
After Fourier transformation, Eqs.~(\ref{eq:dndnxH8}) and (\ref{eq:dndnqSlope2}) imply that
\begin{equation}
\Delta[\Ham_8] = 4g_{\rho+},
\label{eq:DeltaH8grhoplus}
\end{equation}
which is the desired result.
Note that $g_{\rho+}$ is not generally a genuine Luttinger parameter due to
the coupling between the $\rho+$ and $\rho-$ sectors in the C2S2 phase, but should instead be viewed as a
direct measurement of $\Delta[\Ham_8]$ through the density structure factor.

\begin{figure}[t]
\includegraphics[width=1.0\columnwidth]{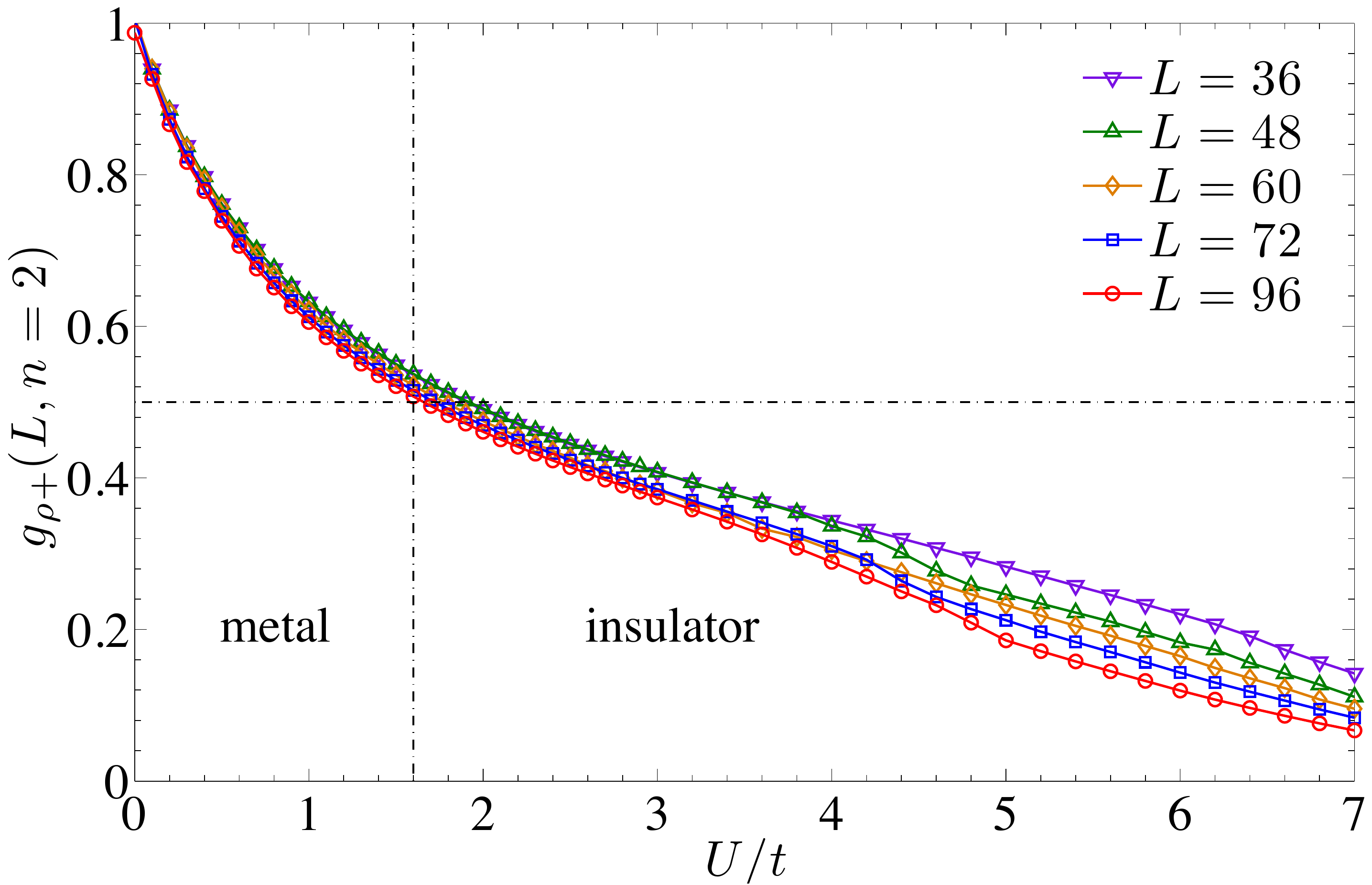}
\caption{
Finite-size estimates of $g_{\rho+}$ [see Eq.~(\ref{eq:grhoplusLn})] versus $U/t$.
Our bosonized theory predicts that measured values of $g_{\rho+}<1/2$ must necessarily
correspond to (flow to) $g_{\rho+}\rightarrow0$ in the thermodynamic limit ($L\rightarrow\infty$).\
The somewhat irregular finite-size behavior in the gapless regions ($U/t\lesssim5.0$) is likely due to ``shell filling'' effects,
i.e., the thermodynamic phase is more readily accommodated by some sizes and less by others.
}
\label{fig:grhoplus_V4}
\end{figure}

In the main text, we relied heavily upon Eq.~(\ref{eq:DeltaH8grhoplus})
to distinguish between metallic and insulating behavior,
where measured $g_{\rho+} > 1/2$ ($g_{\rho+} < 1/2$) implies that $\Ham_8$ is irrelevant (relevant)
so that the system is metallic (insulating).
Of course, if $\Delta[\Ham_8]<2$, the system is necessarily insulating and Eq.~(\ref{eq:dndnqSlope2}) no longer applies;
instead we have $\langle\delta n_{q}\delta n_{-q}\rangle \sim q^2$ as $q\rightarrow0$.
That is, measured $g_{\rho+} < 1/2$ via Eq.~(\ref{eq:dndnqSlope2}) on a finite-size system corresponds
in the thermodynamic limit to $g_{\rho+}\rightarrow0$.  In Fig.~\ref{fig:charges},
even well into the insulating phase of our model as determined by the above arguments,
we see on our $L=96$ site system that apparently $\langle\delta n_{q}\delta n_{-q}\rangle \sim |q|$;
however, with $\Ham_8$ relevant, this must be a finite-size effect due to
the large charge correlation length present in our weak Mott insulating C1S2.

In Fig.~\ref{fig:grhoplus_V4}, we show finite-size estimates of the quantity $g_{\rho+}$
obtained with DMRG for the same parameters of the extended Hubbard model used in the main text.
Specifically, we define
\begin{equation}
g_{\rho+}(L,n) \equiv \frac{L}{4n}\langle\delta n_{q}\delta n_{-q}\rangle\big{|}_{q=n\frac{2\pi}{L}},
\label{eq:grhoplusLn}
\end{equation}
and monitor $g_{\rho+}(L, n=2)$ while varying $U/t$.  This data looks rather far removed from ordinary
KT behavior potentially indicating strong finite-size effects (see also discussion at the end of the previous section).
Still, based on the above analysis, we must have a Mott transition near $U/t=1.6$.
Eventual $g_{\rho+}\rightarrow0$ is expected for all $U/t\gtrsim1.6$,
although that is not apparent on these sizes until deep in the insulating phase, say $U/t\gtrsim4.0$.
We note that the fully gapped C0S0 state ($U/t\gtrsim5.0$) does show clear
$\langle\delta n_{q}\delta n_{-q}\rangle\sim q^2$ behavior, which is not surprising
given the short charge correlation length expected in that state.

\subsubsection{Bosonized representation of operators at finite wavevectors} \label{sec:bosonized_ops}

We now give the bosonized expressions for the spin and density operators at finite ``$2k_F$'' wavevectors
and mathematically establish the Amperean enhancement mechanism summarized in the main text.
Expanding the spin operator as $\mathbf{S}(x)=\sum_Q \mathbf{S}_Q e^{iQx}$,
we can easily write the slowly varying part of the spin operator at various wavevectors, i.e., $\mathbf{S}_Q=\mathbf{S}_Q(x)$,
in terms of the right and left moving electron operators defined in Appendix~\ref{sec:longwave}:
\begin{align}
\mathbf{S}_{2k_{Fa}} &= \frac{1}{2}c^\dagger_{La\alpha}\bm{\sigma}_{\alpha\beta}c_{Ra\beta}, \\
\mathbf{S}_{\pi/2} &= \frac{1}{2}c^\dagger_{R1\alpha}\bm{\sigma}_{\alpha\beta}c_{L2\beta} + \frac{1}{2}c^\dagger_{R2\alpha}\bm{\sigma}_{\alpha\beta}c_{L1\beta}, \\
\mathbf{S}_{k_{F2}-k_{F1}} &= \frac{1}{2}c^\dagger_{R1\alpha}\bm{\sigma}_{\alpha\beta}c_{R2\beta} + \frac{1}{2}c^\dagger_{L2\alpha}\bm{\sigma}_{\alpha\beta}c_{L1\beta}.
\end{align}
Similarly, for the density operator, we have
\begin{align}
\delta n_{2k_{Fa}} &= c^\dagger_{La\alpha}c_{Ra\alpha}, \\
\delta n_{\pi/2} &= c^\dagger_{R1\alpha}c_{L2\alpha} + c^\dagger_{R2\alpha}c_{L1\alpha}, \\
\delta n_{k_{F2}-k_{F1}} &= c^\dagger_{R1\alpha}c_{R2\alpha} + c^\dagger_{L2\alpha}c_{L1\alpha}.
\end{align}
In each case, summations over spin indices are implied,
and $\mathbf{S}_{-Q}=\mathbf{S}^\dagger_{Q}$ and $\delta n_{-Q}=\delta n^\dagger_{Q}$.
Throughout, our use of denoting wavevectors with either $Q$ or $q$ is an attempt to distinguish
the long-wavelength component of an operator, $O_Q$,
from the actual exact operator used in the DMRG, $O_q$.

Bosonizing the above electron bilinears using Eq.~(\ref{eq:bosonization})
results in the following expressions for the spin:
\begin{align}
S^x_{2k_{Fa}} &= -i \eta_{a\uparrow}\eta_{a\downarrow} e^{i\theta_{\rho+}}e^{\pm i\theta_{\rho-}}\sin(\sqrt{2}\varphi_{a\sigma}), \label{eq:Sx2kF} \\
S^y_{2k_{Fa}} &= -i \eta_{a\up} \eta_{a\dn} e^{i\theta_{\rho+}} e^{\pm i\theta_{\rho-}} \cos(\sqrt{2} \varphi_{a\sigma}), \\
S^z_{2k_{Fa}} &= -e^{i\theta_{\rho+}} e^{\pm i\theta_{\rho-}} \sin(\sqrt{2} \theta_{a\sigma}), \label{eq:Sz2kF} \\
S^x_{\pi/2} &= e^{-i\theta_{\rho+}} \Big[
-i \eta_{1\up} \eta_{2\dn} e^{-i \theta_{\sigma-}}
\sin(\varphi_{\rho-} + \varphi_{\sigma+}) \nonumber \\
&~~~~~~~~~~~
-i \eta_{1\dn} \eta_{2\up} e^{i \theta_{\sigma-}}
\sin(\varphi_{\rho-} - \varphi_{\sigma+}) 
\Big], \\
S^y_{\pi/2} &= e^{-i\theta_{\rho+}} \Big[
-i \eta_{1\up} \eta_{2\dn} e^{-i \theta_{\sigma-}}
\cos(\varphi_{\rho-} + \varphi_{\sigma+}) \nonumber \\
&~~~~~~~~~~
+i \eta_{1\dn} \eta_{2\up} e^{i \theta_{\sigma-}}
\cos(\varphi_{\rho-} - \varphi_{\sigma+})
\Big], \\
S^z_{\pi/2} &= e^{-i\theta_{\rho+}} \Big[ 
-i \eta_{1\up} \eta_{2\up} e^{-i \theta_{\sigma+}}
\sin(\varphi_{\rho-} + \varphi_{\sigma-}) \nonumber \\
&~~~~~~~~~~~
+i \eta_{1\dn} \eta_{2\dn} e^{i \theta_{\sigma+}}
\sin(\varphi_{\rho-} - \varphi_{\sigma-})
\Big],
\end{align}
\begin{align}
S^x_{k_{F2}-k_{F1}} &= e^{-i\theta_{\rho-}} \Big[ 
-i \eta_{1\up} \eta_{2\dn} e^{-i \theta_{\sigma+}}
\sin(\varphi_{\rho-} + \varphi_{\sigma+}) \nonumber \\
&~~~~~~~~~~~
-i \eta_{1\dn} \eta_{2\up} e^{i \theta_{\sigma+}}
\sin(\varphi_{\rho-} - \varphi_{\sigma+})
\Big], \\
S^y_{k_{F2}-k_{F1}} &= e^{-i\theta_{\rho-}} \Big[ 
-i \eta_{1\up} \eta_{2\dn} e^{-i \theta_{\sigma+}}
\cos(\varphi_{\rho-} + \varphi_{\sigma+}) \nonumber \\
&~~~~~~~~~~
+i \eta_{1\dn} \eta_{2\up} e^{i \theta_{\sigma+}}
\cos(\varphi_{\rho-} - \varphi_{\sigma+})
\Big], \\
S^z_{k_{F2}-k_{F1}} &= e^{-i\theta_{\rho-}} \Big[ 
-i \eta_{1\up} \eta_{2\up} e^{-i \theta_{\sigma-}}
\sin(\varphi_{\rho-} + \varphi_{\sigma-}) \nonumber \\
&~~~~~~~~~~
+i \eta_{1\dn} \eta_{2\dn} e^{i \theta_{\sigma-}}
\sin(\varphi_{\rho-} - \varphi_{\sigma-})
\Big],
\end{align}
and for the density:
\begin{align}
\delta n_{2k_{Fa}} &= 2i e^{i\theta_{\rho+}} e^{\pm i \theta_{\rho-}} 
\cos(\sqrt{2} \theta_{a\sigma}), \label{eq:deltan2kF} \\
\delta n_{\pi/2} &= 2e^{-i\theta_{\rho+}} \Big[
-i \eta_{1\up} \eta_{2\up} e^{-i\theta_{\sigma+}} 
\sin(\varphi_{\rho-} + \varphi_{\sigma-}) \nonumber \\
&~~~~~~~~~
-i \eta_{1\dn} \eta_{2\dn} e^{i\theta_{\sigma+}} 
\sin(\varphi_{\rho-} - \varphi_{\sigma-}) 
\Big], \label{eq:deltanpiby2} \\
\delta n_{k_{F2}-k_{F1}} &= 2e^{-i\theta_{\rho-}} \Big[
-i \eta_{1\up} \eta_{2\up} e^{-i\theta_{\sigma-}} 
\sin(\varphi_{\rho-} + \varphi_{\sigma-}) \nonumber \\
&~~~~~~~~~
-i \eta_{1\dn} \eta_{2\dn} e^{i\theta_{\sigma-}} 
\sin(\varphi_{\rho-} - \varphi_{\sigma-}) 
\Big],
\end{align}
where for expressions with $\pm$ in the exponent, $+$ refers to band $a=1$, while $-$ refers to band $a=2$.

Perhaps the most important point to take away is that all operators at $Q=2k_{Fa},\pi/2$
are proportional to $e^{\pm i\theta_{\rho+}}$.  Therefore, the fluctuating field content of these
operators is reduced upon gapping out (pinning of) $\theta_{\rho+}$ when crossing the Mott transition
from the C2S2 metal to C1S2 insulator.  This leads to lowering of the associated scaling dimensions
and subsequent enhancement of the structure factor singularities.
To illustrate this concretely, assume for the moment that the $\rho+$ and $\rho-$ sectors are decoupled
in the charge sector Lagrangian for the C2S2, i.e., $A_{12}=A_{21}=B_{12}=B_{21}=0$ in Eq.~(\ref{eq:LrhoFullHami}),
with corresponding Luttinger parameters $g_{\rho+}$ and $g_{\rho-}$.
We then have the following for the scaling dimensions of the above operators:
\begin{equation}
\Delta[\mathbf{S}_{2k_{Fa}}] = \Delta[\delta n_{2k_{Fa}}] =  \frac{1}{2} + \frac{g_{\rho-}}{4} + \frac{g_{\rho+}}{4},
\label{eq:dims_2kF}
\end{equation}
\begin{equation}
\Delta[\mathbf{S}_{\pi/2}] = \Delta[\delta n_{\pi/2}] = \frac{1}{2} + \frac{1}{4g_{\rho-}} + \frac{g_{\rho+}}{4},
\label{eq:dims_piby2}
\end{equation}
\begin{equation}
\Delta[\mathbf{S}_{k_{F2}-k_{F1}}] = \Delta[\delta n_{k_{F2}-k_{F1}}] = \frac{1}{2} + \frac{1}{4g_{\rho-}} + \frac{g_{\rho-}}{4},
\label{eq:dims_kF2mkF1}
\end{equation}
where we have assumed SU(2) invariance, $g_{1\sigma}=g_{2\sigma}=1$ (see the next section).
Right at the Mott transition $g_{\rho+}=1/2$, while immediately on the insulating side $g_{\rho+}\rightarrow0$.
Therefore, the dimensions in Eqs.~(\ref{eq:dims_2kF}) and (\ref{eq:dims_piby2})
corresponding to operators at $Q=2k_{Fa},\pi/2$
should indeed \emph{decrease} at the transition (by an amount of 1/8 in the decoupled approximation).
Such an enhancement of the associated spin structure factor singularities on the insulating side
of the Mott transition is in fact dramatically seen in the DMRG data of Fig.~\ref{fig:spins}.

Furthermore, stability of the C1S2 insulator requires $g_{\rho-}<1$ (see Ref.~\onlinecite{Sheng:2009up}),
which implies $\Delta[\mathbf{S}_{\pi/2}] > \Delta[\mathbf{S}_{2k_{Fa}}]$ (and similarly for $\delta n_Q$).
Thus, for the structure factors in the C1S2 phase, the features at $q=2k_{Fa}$
should be more pronounced than those at $q=\pi/2$.
Indeed, this is observed in the spin structure factor data of Fig.~\ref{fig:spins} on the insulating side
of the Mott transition in our model.  More generally, the presence of clear power-law singularities
in $\langle\mathbf{S}_{q}\cdot\mathbf{S}_{-q}\rangle$ at finite wavevectors
in both the metal and weak Mott insulator points strongly towards to presence of gapless spin excitations
in both phases (see also Appendix~\ref{sec:gsigmaplus}).

Note that the density operator at $Q=2k_{Fa},\pi/2, k_{F2}-k_{F1}$ still remains power law when $\theta_{\rho+}$
gets pinned, i.e., $\delta n_Q$ does not contain the wildly fluctuating field $\varphi_{\rho+}$.
In fact, for $Q=2k_{Fa},\pi/2$ the density also contains directly $\theta_{\rho+}$
[see Eqs.~(\ref{eq:deltan2kF}), (\ref{eq:deltanpiby2})]
%, i.e., $\delta n_Q\sim e^{\pm i\theta_{\rho+}}(\cdots)$
and has the same scaling dimension as the spin operator: $\Delta[\delta n_Q] = \Delta[\mathbf{S}_Q]$!
Therefore, such Friedel oscillations should actually be enhanced in the Mott insulator~\cite{Mross11_PRB_84_041102}.
This enhancement is difficult to see in the density structure factor DMRG data of Fig.~\ref{fig:charges},
but that is likely due to the small amplitudes of the features.
The power-law nature, however, is still apparent, at least around $q=2k_{F1},k_{F2}-k_{F1}$.

The bilinears that get enhanced, i.e., those at $Q=2k_{Fa},\pi/2$, can be predicted by simple ``Amperean rules''.
Specifically, in the (1+1)D U(1) gauge theory formulation of the C1S2 spinon metal phase~\cite{Sheng:2009up},
$\theta_{\rho+}$ corresponds to the mode that is pinned upon inclusion of gauge fluctuations
which implements at long wavelengths the constraint of one spinon per site
(in this language, the up and down spinons carry the same gauge charge).
We then expect that the bilinears that get enhanced upon introducing the gauge fluctuations
are those composed from operators that produce parallel gauge currents,
so-called Amperean attraction~\cite{Lee:2006de, Sheng:2009up}.
This is indeed the case for the spin and density operators at $Q=2k_{Fa},\pi/2$ which involve
a particle and hole moving in opposite directions.  In contrast, the bilinears at $Q=k_{F2}-k_{F1}$
involve operators with antiparallel gauge currents and are therefore not enhanced;
indeed these operators do not contain $\theta_{\rho+}$ at all.
We remark that in our electronic model, the above ``gauge constraint'' is implemented dynamically
by electron repulsion upon pinning of the overall conducting charge mode $\theta_{\rho+}$.

In the main text, we have also used the dimer correlations, as defined and detailed
in Appendix~\ref{sec:observables_def}, to characterize the ground state.  Following Ref.~\onlinecite{Sheng:2009up},
we can approximate the bond energy as the electron hopping energy, i.e.,
$\mathcal{B}(x) \sim -t\sum_\alpha\left[c^\dagger_\alpha(x)c_\alpha(x+1) + \mathrm{H.c.} \right]$.
In fact, in our DMRG measurements it would have been reasonable to use this as the definition
of $\mathcal{B}(x)$, but we instead implemented the full $\mathcal{B}(x)=\mathbf{S}(x)\cdot\mathbf{S}(x+1)$,
which makes the two-point function $\langle\mathcal{B}(x)\mathcal{B}(x')\rangle$
a four-spin (eight-electron) measurement.  In any case, expansion in continuum fields reveals
\begin{equation}
\mathcal{B}_Q \sim e^{iQ/2} \delta n_Q,
\end{equation}
which holds for all $Q\neq\pi$.  Hence, we expect features at the same wavevectors
in measurements of both $\langle\delta n_{q}\delta n_{-q}\rangle$ and $\langle \mathcal{B}_q\mathcal{B}_{-q}\rangle$.
This is indeed observed in Figs.~\ref{fig:charges} and \ref{fig:dimers}, where in the putative C1S2 insulator
the power-law nature of the features is, as expected, much more apparent in the dimer correlations than in the density correlations.

We further note that $\langle \mathcal{B}_q\mathcal{B}_{-q}\rangle$
very clearly picks up a feature at $q=4k_{F2}=-4k_{F1}$,
while this feature is much weaker, though still present, in $\langle\delta n_{q}\delta n_{-q}\rangle$.
As mentioned in the main text, the wavevector $4k_{F2}=-4k_{F1}$ is a four-fermion contribution to the density/bond energy.
Specifically,
\begin{align}
\delta n_{4k_{F1}} &:~c^\dagger_{L1\up}c^\dagger_{L1\dn}c_{R1\up}c_{R1\dn} \sim e^{i2\theta_{\rho+}} e^{i2\theta_{\rho-}}, \\
\delta n_{-4k_{F2}} &:~c^\dagger_{R2\up}c^\dagger_{R2\dn}c_{L2\up}c_{L2\dn} \sim e^{-i2\theta_{\rho+}} e^{i2\theta_{\rho-}},
\end{align}
both contribute with independent numerical prefactors,
and have scaling dimensions in the decoupled $\rho\pm$ approximation of
\begin{equation}
\Delta[\delta n_{4k_{F2}}] = \Delta[\mathcal{B}_{4k_{F2}}] = g_{\rho+} + g_{\rho-}.
\end{equation}
In the C1S2, $g_{\rho+}\rightarrow 0$ so that
$\Delta[\mathcal{B}_{4k_{F2}}] = g_{\rho-}$.  Gaplessness of the spin sector
requires $g_{\rho-} < 1$ (see Refs.~\onlinecite{Sheng:2009up, Motrunich10_PRB_81_045105}).
Hence, the singularity at $q=4k_{F2}$ in $\langle \mathcal{B}_q\mathcal{B}_{-q}\rangle$
should be stronger than a slope discontinuity (unit scaling dimension of the associated operator)---this
indeed appears to be the case in our dimer structure factor data of Fig.~\ref{fig:dimers}.

There is yet another important four-fermion contribution to the spin and density/bond energy
at wavevector $Q=\pi$.  We here focus on the latter, where for the bond energy
we get contributions such as~\cite{Sheng:2009up}
$\mathcal{B}_\pi:~i\delta n_{2k_{F1}}\delta n_{2k_{F2}} + \mathrm{H.c.}$,
which when bosonized gives
\begin{equation}
\mathcal{B}_\pi \sim [\cos(2\theta_{\sigma+}) + \cos(2\theta_{\sigma-})]\sin(2\theta_{\rho+}) + \cdots.
\label{eq:Bpi}
\end{equation}
This operator has unit scaling dimension at the C1S2 fixed point ($\Delta[\mathcal{B}_\pi]=1$)
and should thus correspond to a slope discontinuity in $\langle \mathcal{B}_q\mathcal{B}_{-q}\rangle$ at $q=\pi$.
Remarkably, this appears to be consistent with e.g.~our characteristic C1S2 data point at $U/t=4.0$
as presented in the main text (see curve with green squares in Fig.~\ref{fig:dimers}).
Furthermore, inspecting Eq.~(\ref{eq:Bpi}) reveals that this feature will only be present in the C1S2
if the pinning of $\theta_{\rho+}$ due to relevance of $\Ham_8=2u\cos(4\theta_{\rho+})$
is such that $\sin(2\theta_{\rho+})\neq0$.
This is precisely what we would expect if the pinned value of $\theta_{\rho+}$
occurs at $4\theta_{\rho+} = \pi \mod 2\pi$, which corresponds to the \emph{minima} of $\cos(4\theta_{\rho+})$.
We thus conclude that $u>0$ in our eight-fermion umklapp interaction,
as might initially be expected for repulsively interacting electrons~\cite{Sheng:2009up}.
On the other hand, $u<0$ would lead to pinning of $\theta_{\rho+}$
such that $4\theta_{\rho+} = 0 \mod 2\pi$, i.e., $\sin(2\theta_{\rho+})=0$,
thus killing the feature in $\langle \mathcal{B}_q\mathcal{B}_{-q}\rangle$ at $q=\pi$.

At wavevector $Q=\pi$, the bond-centered density $\mathcal{B}_\pi$ is odd under mirror symmetry ($x\rightarrow-x$),
while the site-centered density $\delta n_\pi$ is even.  Contributions to the latter include
$\delta n_\pi:~\delta n_{2k_{F1}}\delta n_{2k_{F2}} + \mathrm{H.c.}$,
which in terms of the bosonized fields reads
\begin{equation}
\delta n_\pi \sim [\cos(2\theta_{\sigma+}) + \cos(2\theta_{\sigma-})]\cos(2\theta_{\rho+}) + \cdots.
\label{eq:npi}
\end{equation}
Hence, the pinning condition $4\theta_{\rho+}=\pi\mod2\pi$ inferred above implies $\cos(2\theta_{\rho+})=0$.
Indeed, the DMRG data shows no feature in $\langle \delta n_q\delta n_{-q}\rangle$ at $q=\pi$
within the putative C1S2 phase (see Fig.~\ref{fig:charges}).  Again, we conclude that for our system
with repulsively interacting electrons, we must have $u>0$ in $\Ham_8$.

Finally, presence of a feature at $q=\pi$ in $\langle \delta n_q\delta n_{-q}\rangle$ in the C1S2 weak Mott insulator
would lead to long-range period-2 (site-centered) charge density wave order
in the C0S0 strong Mott insulator at very large $U/t$.
This is indeed very unnatural in our model where the on-site $U$ term is the largest interaction energy scale
in the Hamiltonian.  Instead, the strong Mott insulator realized in our model develops period-2 long-range order
in the \emph{bond-centered} density, as evidenced by the Bragg peak
in $\langle \mathcal{B}_q\mathcal{B}_{-q}\rangle$ at $q=\pi$.
The power-law feature at the same wavevector in the weak Mott insulator [see Eq.~(\ref{eq:Bpi})]
is the precursor of this eventual long-range VBS order at large $U/t$.

We finally discuss the electron operator itself [Eq.~(\ref{eq:bosonization})],
which is of course the most primitive operator of all.
When written in terms of ``$\rho\pm$'' and ``$a\sigma$'' modes, we have
\begin{align}
c_{Pa\alpha} = \eta_{a\alpha} \exp&\left\{\frac{i}{\sqrt{2}}\left[\frac{1}{\sqrt{2}}(\varphi_{\rho+}\pm\varphi_{\rho-})\pm\varphi_{a\sigma}\right] \right. \nonumber \\ 
&\left. + \frac{iP}{\sqrt{2}}\left[\frac{1}{\sqrt{2}}(\theta_{\rho+}\pm\theta_{\rho-})\pm\theta_{a\sigma}\right] \right\},
\label{eq:electronExplicit}
\end{align}
where the first $\pm$ on each line refers to $a=1,2$, while the second refers to $\alpha=\,\up,\dn$.
Of course, once the $\theta_{\rho+}$ field is pinned,
the electron Green's function $\langle c^\dagger_\alpha(x)c_\alpha(0)\rangle$
is expected to decay exponentially at all wavevectors.
Mathematically, this is due to its conjugate field $\varphi_{\rho+}$ also being present in the bosonized
representation of the electron operator:  By the uncertainty principle, pinning of $\theta_{\rho+}$
will cause $\varphi_{\rho+}$ to fluctuate wildly leading to exponential decay of the Green's function.
While it is somewhat difficult to ascertain this exponential decay within the putative C1S2 phase
for the electron momentum distribution function DMRG data of Fig.~\ref{fig:electron}, we again believe this is due
to the excessively large charge correlation lengths present in our electronic spinon metal.

From Eq.~(\ref{eq:electronExplicit}), we also see that gapping of a spin mode will cause the associated electron
Fermi point to gap out, and thus the electron Green's function can in principle detect spin-gap behavior.
However, this is rather difficult in practice~\cite{Japaridze07_PRB_76_115118}, and in the following section
we discuss a better approach as employed in the main text.

\subsubsection{Assessing gaplessness of the spin sector through $g_{\sigma+}$} \label{sec:gsigmaplus}

Inspection of the bosonized expressions for the different components of the spin operator
at wavevectors $Q=2k_{Fa}$ in Eqs.~(\ref{eq:Sx2kF})-(\ref{eq:Sz2kF}), reveals that in the fixed-point theory
for either the C2S2 metal or C1S2 insulator we must have only trivial Luttinger parameters in the spin sector:
$g_{1\sigma}=g_{2\sigma}=1$.  Specifically, for arbitrary $g_{a\sigma}$ as in Eq.~(\ref{eq:spinHami})
and decoupled $\rho+$ and $\rho-$ modes as in the illustrative discussion in Appendix~\ref{sec:bosonized_ops} above,
we have
\begin{equation}
\Delta[S^x_{2k_{Fa}}] = \Delta[S^y_{2k_{Fa}}] = \frac{g_{\rho+}}{4} + \frac{g_{\rho-}}{4} + \frac{1}{2g_{a\sigma}},
\end{equation}
\begin{equation}
\Delta[S^z_{2k_{Fa}}] = \frac{g_{\rho+}}{4} + \frac{g_{\rho-}}{4} + \frac{g_{a\sigma}}{2},
\end{equation}
where in the C1S2 insulator we have $g_{\rho+}\rightarrow0$.
Therefore, SU(2) spin invariance manifest through isotropic spin-spin correlations functions at wavevectors
$2k_{Fa}$, i.e., $\Delta[S^x_{2k_{Fa}}]  = \Delta[S^y_{2k_{Fa}}] = \Delta[S^z_{2k_{Fa}}]$, 
indeed dictates that
\begin{equation}
g_{1\sigma}=g_{2\sigma}=1,
\label{eq:gsigma_condition}
\end{equation}
which constitutes a simple generalization of the well-known one-mode case~\cite{Giamarchi}
(see also Ref.~\onlinecite{Sedlmayr13_PRB_88_195113}).

We now show how measurement of the spin structure factor at zero momentum can assess
the condition in Eq.~(\ref{eq:gsigma_condition}).  The slowly varying part of the spin density is
$S^z(x) = \partial_x\theta_{\sigma+}/\pi$, hence the long-wavelength part of the real-space spin-spin
correlation function evaluated in the fixed-point theory for either the C2S2 or C1S2
[see Eq.~(\ref{eq:spinHami})] reads
\begin{equation}
\langle S^z(x) S^z(0) \rangle = -\frac{g_{\sigma+}}{2\pi^2}\frac{1}{x^2} + \cdots,
\label{eq:SzSzx}
\end{equation}
where we have defined
\begin{equation}
g_{\sigma+} \equiv \frac{g_{1\sigma} + g_{2\sigma}}{2}.
\end{equation}
Equation (\ref{eq:SzSzx}) gives for the spin structure factor as $q\rightarrow0$:
\begin{equation}
\langle S^z_{q} S^z_{-q} \rangle = \frac{g_{\sigma+}}{2\pi}|q|,
\end{equation}
which we use in the main text to estimate the parameter $g_{\sigma+}$
[see Eq.~(\ref{eq:sdotsqSlope}) and the inset of Fig.~\ref{fig:spins}].
Clearly then within the fixed-point theory we should have $g_{\sigma+}=1$,
while in the presence of a spin gap $\langle S^z_{q}S^z_{-q}\rangle \sim q^2$,
so that $g_{\sigma+}\rightarrow0$.  Note that, as with $g_{\rho+}$ above, $g_{\sigma+}$
is not a genuine Luttinger parameter as even free electrons are not generally
diagonal in the $\sigma\pm$ basis.

The above considerations are valid for the fixed point in the thermodynamic limit.
However, there are several marginal interactions that need to be irrelevant
for the spin sector to remain gapless and the C2S2 and C1S2 to be stable phases.
Thus, the presence of such marginally irrelevant interactions
will affect measurement of $g_{\sigma+}$ on finite-size systems.
In the case of our C2S2 and C1S2, the residual interactions in the spin sector
that mix right and left movers read
\begin{equation}
\Ham^\sigma_{RL} = -\sum_{a,b}
\left( w^\sigma_{ab} \mathbf{J}_{R a b} \cdot \mathbf{J}_{L a b} 
       + \lambda^\sigma_{ab} \mathbf{J}_{R a a} \cdot \mathbf{J}_{L b b}
\right),
\end{equation}
where $\mathbf{J}_{P a b} \equiv \frac{1}{2}c^\dagger_{Pa\alpha}\bm{\sigma}_{\alpha\beta}c_{Pb\beta}$.
In the C2S2 and C1S2, the $w^\sigma_{ab}$ terms are strictly irrelevant,
while the $\lambda^\sigma_{ab}$ terms are only marginally irrelevant~\cite{Sheng:2009up,Motrunich10_PRB_81_045105}.
Bosonizing the latter interactions gives
\begin{align}
\tilde\Ham^\sigma_{RL} &= V_z + V_\perp, \\
V_z &= \sum_a \frac{\lambda_{aa}^\sigma}{8\pi^2}
\left[(\partial_x \varphi_{a\sigma})^2 - (\partial_x \theta_{a\sigma})^2
\right] \\
&+ \frac{\lambda_{12}^\sigma}{4\pi^2} 
\left[(\partial_x \varphi_{1\sigma}) (\partial_x \varphi_{2\sigma}) 
      - (\partial_x \theta_{1\sigma}) (\partial_x \theta_{2\sigma})
\right], \\
V_\perp &= \sum_a \lambda_{aa}^\sigma \cos(2\sqrt{2}\theta_{a\sigma}) \\
&+ 2\lambda_{12}^\sigma \hat\Gamma \cos(2\theta_{\sigma+}) 
\cos(2\varphi_{\sigma-}),
\end{align}
where $\hat\Gamma \equiv \eta_{1\up} \eta_{1\dn} \eta_{2\up} \eta_{2\dn}$.

A necessary condition for the spin to be gapless is that the couplings $\lambda_{ab}^\sigma$
be initially positive, corresponding to the system being overall repulsive in the spin sector.
Ultimate stability of the C2S2 and C1S2 corresponds to $\lambda_{ab}^\sigma$
renormalizing to zero via slow marginal flows.  It should in principle be possible to
calculate precise flows (and finite-size scaling behavior) of our effective $g_{\sigma+}$ parameter
by analyzing the behavior of the zero-momentum piece of the spin structure factor
perturbatively in the $\lambda_{ab}^\sigma$.  We do not pursue this here, but instead to get a rough, initial
feel for the trends within our Abelian bosonization, imagine for the moment naively ignoring the $V_\perp$ cosines
and $\lambda_{12}^\sigma$ cross terms.
Then, the quadratic $V_z$ terms effectively feed into renormalizing the $g_{a\sigma}$ Luttinger parameters
above (below) unity for $\lambda_{aa}^\sigma$ positive (negative), hence effectively corresponding to
$g_{\sigma+} > 1$ ($g_{\sigma+} < 1$) on a finite-size system.
This is indeed the expected trend for overall repulsion in the spin sector.

On the other hand, the flows for the C1S0 superconductor (the main instability of the C2S2)
correspond to $\lambda_{aa}^\sigma$ eventually becoming negative (attraction in the spin sector)
and then diverging to $-\infty$.  All modes then eventually get gapped out except the overall conducting
$\rho+$ mode~\cite{Balents96_PRB_53_12133, Gros01_PRB_64_113106}, so that for the spin structure factor we have
$\langle S^z_{q}S^z_{-q}\rangle \sim q^2$ as $q\rightarrow0$, i.e., $g_{\sigma+}\rightarrow0$.
On a finite-size system, we thus expect the spin gap to be manifest as a measured $g_{\sigma+} < 1$.
Note, though, that due to \emph{initial} repulsion in the spin sector [$\lambda_{ab}^\sigma(\ell=0) > 0$],
even an eventual C1S0 may exhibit ``stiffening'' of the spin sector on relatively short length scales,
i.e., measured $g_{\sigma+} > 1$.
These considerations highlight why it is so difficult to detect spin-gap behavior
in models such as the $t$-$t'$-$U$ Hubbard model~\cite{Japaridze07_PRB_76_115118}.
We stress, however, that in our model with longer-ranged repulsion---a model which is known to be spin gapless
at weak coupling ($U/t\ll 1$) for our chosen parameters~\cite{Motrunich10_PRB_81_045105}---measurements of
$g_{\sigma+}$ still strongly indicate spin gaplessness all the way up to $U/t\simeq5.0$,
well past the Mott critical value of $U/t=1.6$.  In the next section, we contrast this with
the behavior of the on-site $t$-$t'$-$U$ Hubbard model at $\kappa=0$ in which the metal
and insulator are presumably both spin gapped.

Finally, we again mention that the observed power-law singularities in the spin structure factor
at the various ``$2k_F$'' wavevectors (see the main text and Appendix~\ref{sec:bosonized_ops})
provide complementary evidence that the spin sector is gapless in both the metal (C2S2)
and weak Mott insulator (C1S2) of our model.

\subsection{Further analysis of $g_{\sigma+}$ DMRG data} \label{sec:gsigmaplus_dmrg}

Here we present more data of our DMRG measurements of the parameter $g_{\sigma+}$
discussed in the previous section.  Specifically, we define a finite-size estimate
of $g_{\sigma+}$ via Eq.~(\ref{eq:sdotsqSlope}) by evaluating the slope
of the spin structure factor at a momentum $q=n\frac{2\pi}{L}$ with $n$ a small integer:
\begin{equation}
g_{\sigma+}(L,n) \equiv \frac{L}{3n}\langle\mathbf{S}_{q}\cdot\mathbf{S}_{-q}\rangle\big{|}_{q=n\frac{2\pi}{L}},
\label{eq:gsigmaplusLn}
\end{equation}
where in what follows we choose $n=2$.

\begin{figure}[t]
\includegraphics[width=1.0\columnwidth]{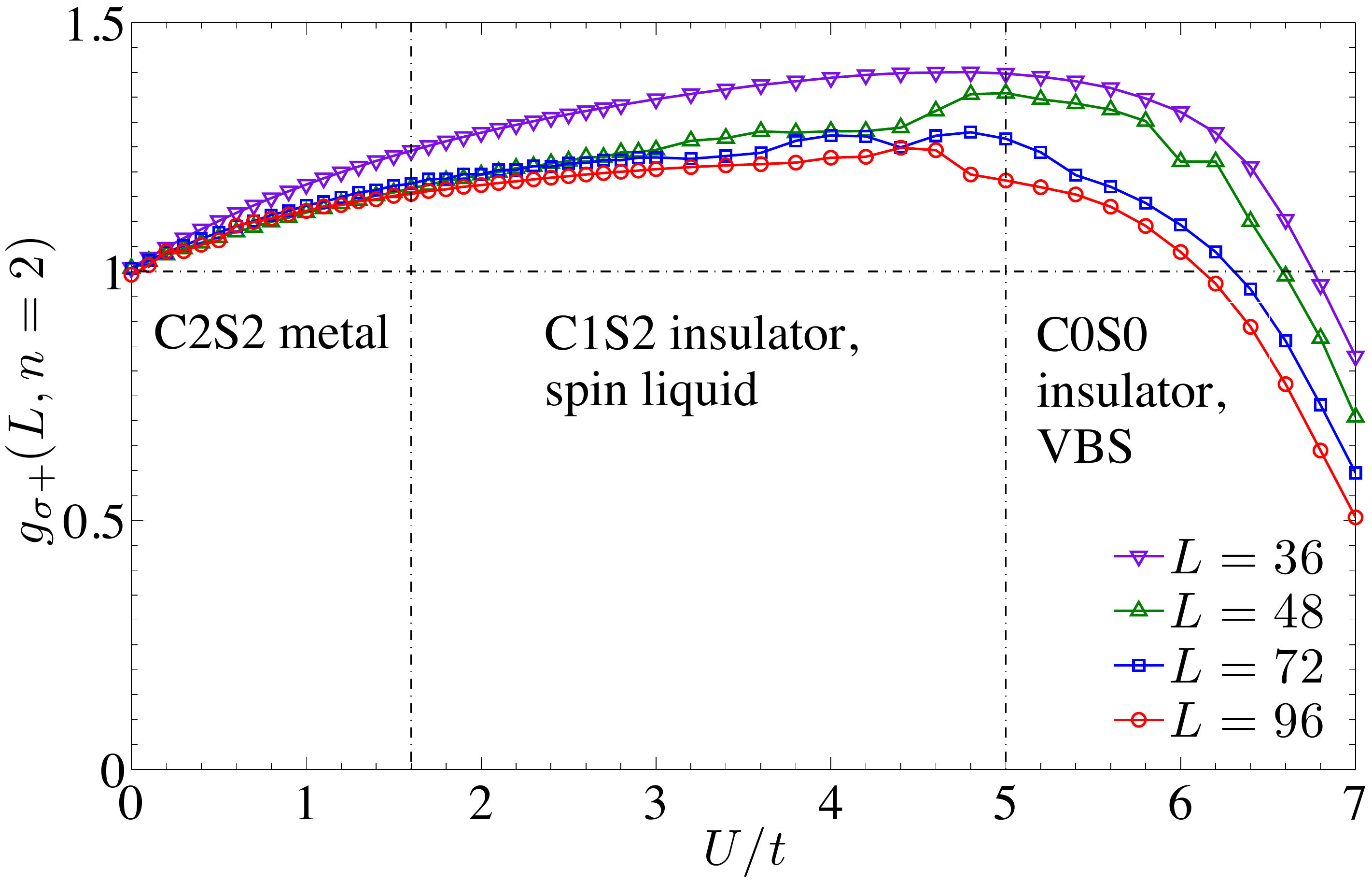}
\caption{
Finite-size estimates of $g_{\sigma+}$ [see Eq.~(\ref{eq:gsigmaplusLn})] versus $U/t$
for the same parameters in the extended Hubbard model at the focus of the main text.
The putative realized phases (see text) are labeled with separating vertical dashed-dotted lines.
At $U/t=0$, our DMRG calculations give $g_{\sigma+}(L,n=2)=1$ to within 1\% for all sizes;
this serves as a very useful check on our convergence since free electrons are, ironically,
very challenging to converge in the DMRG.
}
\label{fig:gsigmaplus_V4}
\end{figure}

\begin{figure}[t]
\includegraphics[width=1.0\columnwidth]{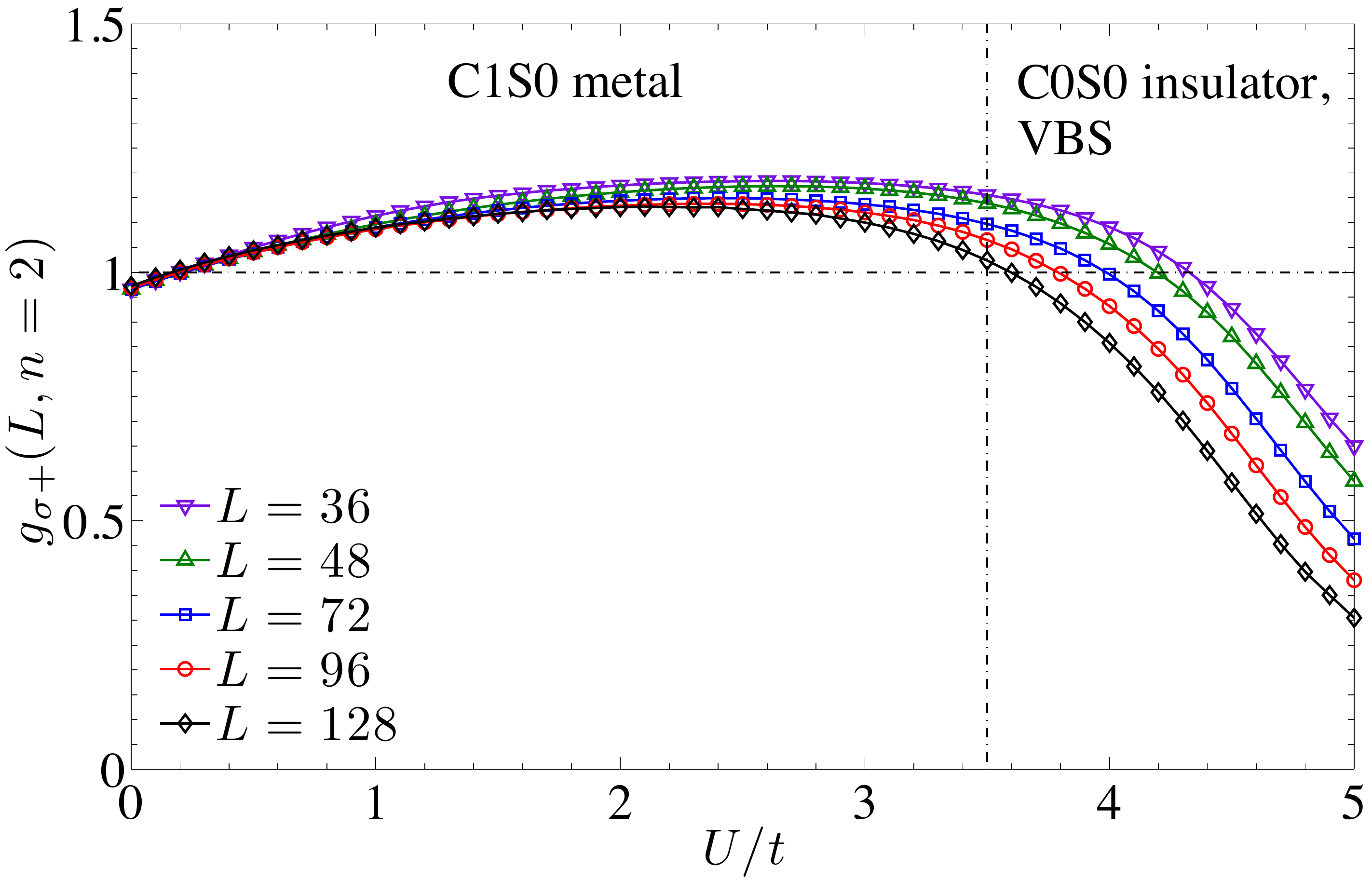}
\caption{
Finite-size estimates of $g_{\sigma+}$ [see Eq.~(\ref{eq:gsigmaplusLn})] versus $U/t$
for the on-site $t$-$t'$-$U$ Hubbard model at $t'/t=0.8$.
%[Eqs.~(\ref{eq:hamiltonian})-(\ref{eq:modelV}) with $\kappa=0$].
The vertical dashed-dotted line at $U/t=3.5$ indicates our estimate of the Mott transition
between the C1S0 metal and C0S0 period-2 VBS insulator from $g_{\rho+}$ measurements
(not shown; see Ref.~\onlinecite{Mishmash_unpublished}).
This value is in good agreement with earlier studies
of the half-filled $t$-$t'$-$U$ Hubbard model~\cite{Japaridze07_PRB_76_115118, Tocchio10_PRB_81_205109}.
Here, we use open boundary conditions which gives very good convergence,
though at the expense of some small systematic error in determining $g_{\sigma+}$
from the momentum-space structure factor;
e.g., $g_{\sigma+}(L,n=2)$ is slightly less than one at $U/t=0$ which is due entirely
to the usage of open boundary conditions.
}
\label{fig:gsigmaplus_Hubbard}
\end{figure}

In Figs.~\ref{fig:gsigmaplus_V4} and \ref{fig:gsigmaplus_Hubbard}, we show $g_{\sigma+}(L,n=2)$
versus $U/t$ on several system sizes $L$ for the extended Hubbard model as presented in the main text
[Eqs.~(\ref{eq:hamiltonian})-(\ref{eq:modelV}) with $t'/t=0.8$, $\kappa=0.5$, $\gamma=0.2$]
and the on-site $t$-$t'$-$U$ Hubbard model
[Eqs.~(\ref{eq:hamiltonian})-(\ref{eq:modelV}) with $t'/t=0.8$, $\kappa=0$], respectively.
In the former case, we use periodic boundary conditions due to the reasons discussed
in Appendix~\ref{sec:observables_def}, while in the latter case we use standard open boundary conditions.
Note that the $L=96$ data in Fig.~\ref{fig:gsigmaplus_V4} corresponds to the second
($q=2\frac{2\pi}{96}$) data points in the inset of Fig.~\ref{fig:spins}.

We first focus on the extended Hubbard model data as shown in Fig.~\ref{fig:gsigmaplus_V4}.
Here, $g_{\sigma+}(L)$ increases above unity as we turn on $U/t$ and continues to do so well
past the putative Mott transition from the C2S2 metal to C1S2 insulator at $U/t=1.6$.
Rather remarkably, the data does not start renormalizing visibly downwards until $U/t\gtrsim4.0$.
Around $U/t\simeq5.0$, the system starts showing signs of spin-gap behavior
(e.g., a Bragg peak in the dimer structure factor; see Fig.~\ref{fig:dimers})
near which $g_{\sigma+}(L)$ finally starts bending downward.
While the data points on the large sizes are still not fully converged due
to the periodic boundary conditions and inherent difficulty involved in converging
such a quantity at small momenta, we believe that as $L\rightarrow\infty$
we would find $g_{\sigma+}=1$ for $U/t\lesssim5.0$ and $g_{\sigma+}=0$ for $U/t\gtrsim5.0$
(see the previous section).

We here mention that we are not generally able to converge perfectly to a spin-singlet in our DMRG simulations.
To assess this, we can measure the total spin $S_\mathrm{tot}$ in the ground state
(we work only in the $S^z_\mathrm{tot}=0$ sector in the DMRG)
by evaluating the computed spin structure factor at $q=0$:
\begin{equation}
\langle\mathbf{S}_{q}\cdot\mathbf{S}_{-q}\rangle\big{|}_{q=0} = \frac{1}{L}\langle\mathbf{S}_\mathrm{tot}^2\rangle
= \frac{1}{L}S_\mathrm{tot}(S_\mathrm{tot} + 1).
\end{equation}
In simulations of Eqs.~(\ref{eq:hamiltonian})-(\ref{eq:modelV}) with periodic boundary conditions,
we often find for $S_\mathrm{tot}$ some small noninteger value on the order of unity.
For example, on $L=96$ sites with $m=6000$ states,
at the free electron point $U/t=0$, we find $S_\mathrm{tot}=0.60$,
and at the characteristic C1S2 spinon metal point $U/t=4.0$, we find $S_\mathrm{tot}=0.46$.
However, we believe this is just a benign effect of our inability to fully converge the DMRG
and the eventual ground state at $m\rightarrow\infty$ will be a spin-singlet with $S_\mathrm{tot}=0$.
We know this to be true at $U/t=0$, while all indications point toward a spin-singlet C1S2 for $1.6<U/t\lesssim5.0$,
e.g., the features at $2k_{F1}$ and $2k_{F2}$ are symmetrically located about $q=\pi/2$
in measurements of $\langle\mathbf{S}_{q}\cdot\mathbf{S}_{-q}\rangle$ (see Fig.~\ref{fig:spins}).
In fact, this convergence difficulty is to be expected in our parameter regime of $t'/t=0.8$,
as realization of the two-band spinon metal in a pure spin model with ring exchanges
(Ref.~\onlinecite{Sheng:2009up}) found similar DMRG convergence problems
in the corresponding parameter regime of that model.

Also, these difficulties are likely responsible for the small ``jumps'' in the data in Fig.~\ref{fig:gsigmaplus_V4},
since measured finite total spin will have a small, somewhat unpredictable, quantitative effect
on our $g_{\sigma+}(L,n)$ values.  For instance, we are able to converge to a singlet for all $U/t$
on the $L=36$ site system, and hence its curve is smooth.
On the other hand, on the $L=48$ site system, the measured total spin starts abruptly dropping toward
zero near $U/t=4.4$, and we believe this behavior is responsible for the corresponding feature
in the $L=48$ curve of Fig.~\ref{fig:gsigmaplus_V4}.  Ultimately, however, these convergence problems
will almost certainly have no qualitative effect on our conclusions being drawn from the $g_{\sigma+}$ data.

In Fig.~\ref{fig:gsigmaplus_Hubbard}, we show analogous $g_{\sigma+}(L,n=2)$ measurements
for the ordinary on-site $t$-$t'$-$U$ Hubbard model at $t'/t=0.8$.
This model has a spin gap at weak coupling $U/t\ll 1$
(see, e.g., Refs.~\onlinecite{Balents96_PRB_53_12133, Gros01_PRB_64_113106})
so that at small finite interaction strengths we expect the system to be in a spin-gapped C1S0 phase.
However, the RG flows which describe the opening of this spin gap are rather intricate.
Specifically, due to the repulsive Hubbard $U$, the system is initially repulsive (stable) in the spin sector,
while the eventual gapping out of both the spin modes and the ``$\rho-$'' mode happens due
to a delicate interplay of all channels (see Fig.~3 of Ref.~\onlinecite{Motrunich10_PRB_81_045105}).
We believe this initial repulsion in the spin sector is responsible for measured $g_{\sigma+} > 1$
(see also discussion in the previous section), while it will drop below unity for large enough sizes.
On the other hand, if the spin sector is initially attractive (unstable), then we observe $g_{\sigma+} < 1$ for all sizes.
This occurs, e.g., in electronic models with explicit Heisenberg coupling $J\mathbf{S}_i\cdot\mathbf{S}_j$
that favors a spin-gapped (Luther-Emery) liquid (see Ref.~\onlinecite{Mishmash_unpublished}).

The Mott transition in the $t$-$t'$-$U$ Hubbard model will also be driven
by the same eight-fermion umklapp term discussed above.
By measuring its scaling dimension in the same fashion as we have done for the extended model
(see Fig.~\ref{fig:charges} and Appendix~\ref{sec:observables_theory}), we have determined that
for the $U$-only Hubbard model at $t'/t=0.8$ the Mott transition occurs near $U/t=3.5$,
after which period-2 VBS order sets in immediately
(see Ref.~\onlinecite{Mishmash_unpublished} for more details).
We see, however, that $g_{\sigma+}(L)$ already starts bending downward well before then.
We stress that this is in sharp contrast to the data of Fig.~\ref{fig:gsigmaplus_V4} in which our model
with longer-ranged repulsion shows no signs of a spin gap until \emph{well past} the Mott transition.
In that case, the intervening phase is the spin gapless C1S2 spin liquid insulator.

\bibliographystyle{prsty}
\bibliography{mott_transition_biblio}

\end{document}